\documentclass[iop]{emulateapj}

\usepackage{hyperref}
\usepackage{cleveref}

\newcommand{\kms}{{\mathrm{\, km \,s^{-1}}}}

\begin{document}
\title{Large-Amplitude Longitudinal Oscillations Triggered by the Merging of Two Solar Filaments: Observations and Magnetic Field Analysis}

\author{M. Luna\altaffilmark{1,2}, Y. Su\altaffilmark{3,4}, B. Schmieder \altaffilmark{5},  R. Chandra\altaffilmark{6,5} \& T. A. Kucera\altaffilmark{7}}

\altaffiltext{1}{Instituto de Astrof\'{\i}sica de Canarias, E-38200 La Laguna, Tenerife, Spain}
\altaffiltext{2}{Departamento de Astrof\'{\i}sica, Universidad de La Laguna, E-38206 La Laguna, Tenerife, Spain}
\altaffiltext{3}{Key Laboratory for Dark Matter and Space Science, Purple Mountain Observatory, CAS, Nanjing 210008, China}
\altaffiltext{4}{School of Astronomy and Space Science, University of Science and Technology of China, Hefei, Anhui 230026, China}
\altaffiltext{5}{Observatoire de Paris, LESIA, UMR8109 (CNRS), 92195 Meudon Principal Cedex, France}
\altaffiltext{6}{Department of Physics, DSB Campus, Kumaun University, Nainital-- 263001, India }
\altaffiltext{7}{NASA Goddard Space Flight Center, Code 671, Greenbelt, MD 20771, USA}

\shorttitle{Large-Amplitude Longitudinal Oscillations Triggered by the Merging of Two Solar Filaments}
\shortauthors{Luna, Su, Schmieder, Chandra \& Kucera}

\begin{abstract}
We follow the eruption of two related intermediate filaments observed in H$\alpha$ (from GONG) and in EUV (from SDO/AIA) and the resulting large-amplitude longitudinal oscillations of the plasma in the filament channels. The events occurred in and around the decayed active region AR12486 on 2016 January 26. Our detailed study of the oscillation reveals that the periods of the oscillations are about one hour. In H$\alpha$ the period decreases with time and exhibits strong damping. The analysis of 171~\AA\ images shows that the oscillation has two phases, an initial long period phase and a subsequent oscillation with a shorter period. In this wavelength the damping appears weaker than in H$\alpha$. The velocity is the largest ever detected in a prominence oscillation, approximately 100 $\kms$. Using SDO/HMI magnetograms we reconstruct the magnetic field of the filaments modeled as flux ropes by using a flux-rope insertion method.  Applying seismological techniques we determine that the radii of curvature of the field lines in which cool plasma is condensed are in the range 75-120~Mm, in agreement with the reconstructed field. In addition, we infer a field strength of $\ge7$ to 30 gauss, depending on the electron density assumed; that is also in agreement with the values from the reconstruction (8-20 gauss). The poloidal flux is zero and the axis flux is of the order of  10$^{20}$ to 10$^{21}$ Mx, confirming the high shear existing even in a non-active filament.
\end{abstract}

\section{Introduction}\label{sec:introduction}
Solar prominences (filaments when observed on the disk) are intriguing structures embedded in the solar corona. They have been intensively observed and studied since the 19th century \citep[see e.g.,][]{tandberg1995,labrosse2010,mackay2010}. It is believed today that the cool dense plasma of prominences is suspended in magnetic flux ropes or sheared arcades due to the upward magnetic tension of the magnetic field \citep{aulanier1998,gunar2013,gunar2016}. Observations reveal that even so-called ``quiescent'' prominences are highly dynamic, consisting of flowing plasma \citep{zirker1994,zirker1998,schmieder2014}, plumes, bubbles \citep{berger2008,berger2010,dudik2012,Heinzel2008a,gunar2014} and ubiquitous oscillations and waves \citep[see e.g.,][]{okamoto2007,schmieder2013,arregui2012}.

Prominence oscillations have been broadly divided into small- and large-amplitude types according to the classification of \citet{oliver2002}. Small-amplitude oscillations ($<2-3\kms$) are highly localized within a small portion of a filament, reflecting only local and small-scale plasma properties. In contrast, large-amplitude oscillations (LAOs) involve motions with velocities above $20\kms$, and large portions of the filament move in phase, reflecting global characteristics of the plasma and field structure. Such motions appear to be triggered by solar energetic events such as distant or nearby flares, jets, and eruptions. These triggers perturb the filaments producing important displacements with respect to the equilibrium position.

Observations reveal that there are two kinds of LAO motions: transverse and longitudinal with respect to the filament spine \citep[see review by][]{tripathi2009}. The excitation mechanisms and probably the underlying physics of both types of oscillations are different. Transverse LAOs are typically excited by energetic disturbances produced by distant flares, Moreton waves and CMEs \citep[see e.g.,][]{moreton1960,hyder1966,ramsey1966,eto2002,okamoto2004,isobe2006,pouget2006,isobe2007,gilbert2008,hershaw2011,liu2013}. In contrast, large-amplitude longitudinal oscillations (LALOs) are excited by nearby impulsive events, e.g., microflares and small jets \citep{jing2003,jing2006,vrsnak2007,chen2008,zhang2012,Li2012,luna2014}.

In all the LALOs reported so far the periods range from 40 to 100 minutes and the amplitudes are between 20 and 92 $\kms$. In LALOs the prominence threads move parallel to their longest extent, indicating that the motion is along the magnetic field. \citet{luna2014} demonstrated that in the event analyzed the motion formed at a 25$^\circ$ angle with respect to the filament spine. That orientation coincided with the typical observed orientations of  filament magnetic fields \citep[see, e.g.,][]{Leroy1983a,Leroy1984a}. Thus, observational evidence suggests that LALOs are actually oscillations along the magnetic field.

Based on 1D numerical simulations, \citet{luna2012b} and \citet{luna2012c} proposed a pendulum model to explain LALOs, where the restoring force is the gravity projected along the magnetic field lines. These results were confirmed by other 1D simulations by \citet{zhang2012,Zhang2013a}. The simulations show that in LALOs the gas pressure gradients are small and do not influence the oscillation, indicating that the nature of LALOs is not magnetosonic. \citet{luna2016a} found that the oscillations are dependent chiefly on the radius of curvature of the magnetic field and not other details of the field line geometry. \citet{luna2016} found that there is no coupling between longitudinal motion and transverse motions in 2D numerical simulations. All these results indicate the robustness of the pendulum model for explanation and analysis of LALOs.

Large-amplitude oscillations offer a new method for estimating the hard-to-measure prominence plasma and magnetic field structure, by combining observations and theoretical modeling through a technique known as large-amplitude prominence seismology. LAO properties such as the period, damping time, and orientation are directly related to the prominence morphology. Using the LALOs model by \cite{luna2012b} the geometry of the dipped field lines that support the prominence against gravity can be inferred. The magnetic field strength can also be determined using the same model. \citet{luna2014} applied this seismological technique to LALOs for the first time. The authors analyzed the oscillations over the entire filament and obtained the geometry of the field lines and the field intensity. \citet{bi2014} studied LALOs during the slow rise of a filament. They found that the period increased with time and that the magnetic fields supporting the filament become flatter during the slow rising phase using LALO seismology. \citet{zhang2017} also used the LALO seismology to infer the geometry of the filament structure and magnetic field intensity.

 In the present paper a LALO occurring in a large filament observed on 2016 January 23 after a partial eruption and merging are presented in Section \ref{sec:observations}. We analyze the filament oscillations observed using H$\alpha$ data from the Global Oscillations Network (GONG) and 171~\AA\ data from Solar Dynamics Observatory (SDO)'s Atmospheric Imaging assembly (AIA) in Section~\ref{sec:analysis_oscillation}. In Section \ref{sec:flux-rope-model}, the filament magnetic structure is reconstructed by using the flux-rope insertion method of \citet{vanballegooijen2004}. In Section \ref{sec:fieldgeometry} we present the seismological analysis of the oscillations. We find very consistent results by comparing the parameters of the filament derived from seismology (radius of the flux rope, magnetic field strength) with the magnetic field modeling results of Section \ref{sec:flux-rope-model}. Our findings are summarized in Section \ref{sec:conclusions}.

\section{Description of the observations: triggering of the oscillations}\label{sec:observations}

The filaments studied were intermediate type filaments \citep{martin98, engvold15} located in the southern part of the decayed NOAA active region 12486. The two filaments, F1 and F2, partially erupted and merged on 2016 January 26.  As reported by \citet{zheng2017}, these eruptions were associated with two CMEs observed by the SOHO/LASCO coronagraph. Using the SDO/Helioseismic and Magnetic Imager (HMI) photospheric magnetic field data, Zheng et al.\  calculated the magnetic flux of the filament environment and found flux cancellation trends in a broad photosphere area. However, a local flux emergence event was identified near one footpoint that was thought to be responsible for the rise of filament F1. 

We focus our study on the longitudinal oscillations of the non-erupting filament (F2) triggered by the eruption/activation of the other filament (F1). For this study, we use the ground-based GONG H$\alpha$ line 
center data and the high spatial and temporal resolution AIA  \citep{Lemen2012a} on-board the \citep[SDO,][]{Pesnell12} satellite. The pixel size and temporal resolution of the H$\alpha$ data were 1 arcsec and 1 minute respectively. The AIA instrument observes the full Sun in different UV and EUV wavebands with a pixel size of 0.6 arcsec and a maximum cadence of 12 s. In addition, for the photospheric magnetic field we used the data from the HMI \citep{Scherrer12}. The cadence of the line-of-sight (LOS) HMI data is 45 s and the pixel size is 0.5 arcsec. 

In order to better spatially align the GONG and AIA data we compared the locations of sunspots in GONG H$\alpha$ and the AIA 1700~\AA\ band. We find that by rolling the H$\alpha$ images 0.5 degrees counter-clockwise we obtain a very good correspondence of sunspots and filaments in both data sets and in the temporal range considered in this work.

\begin{figure*}[!ht]
\centering\includegraphics[width=1\textwidth]{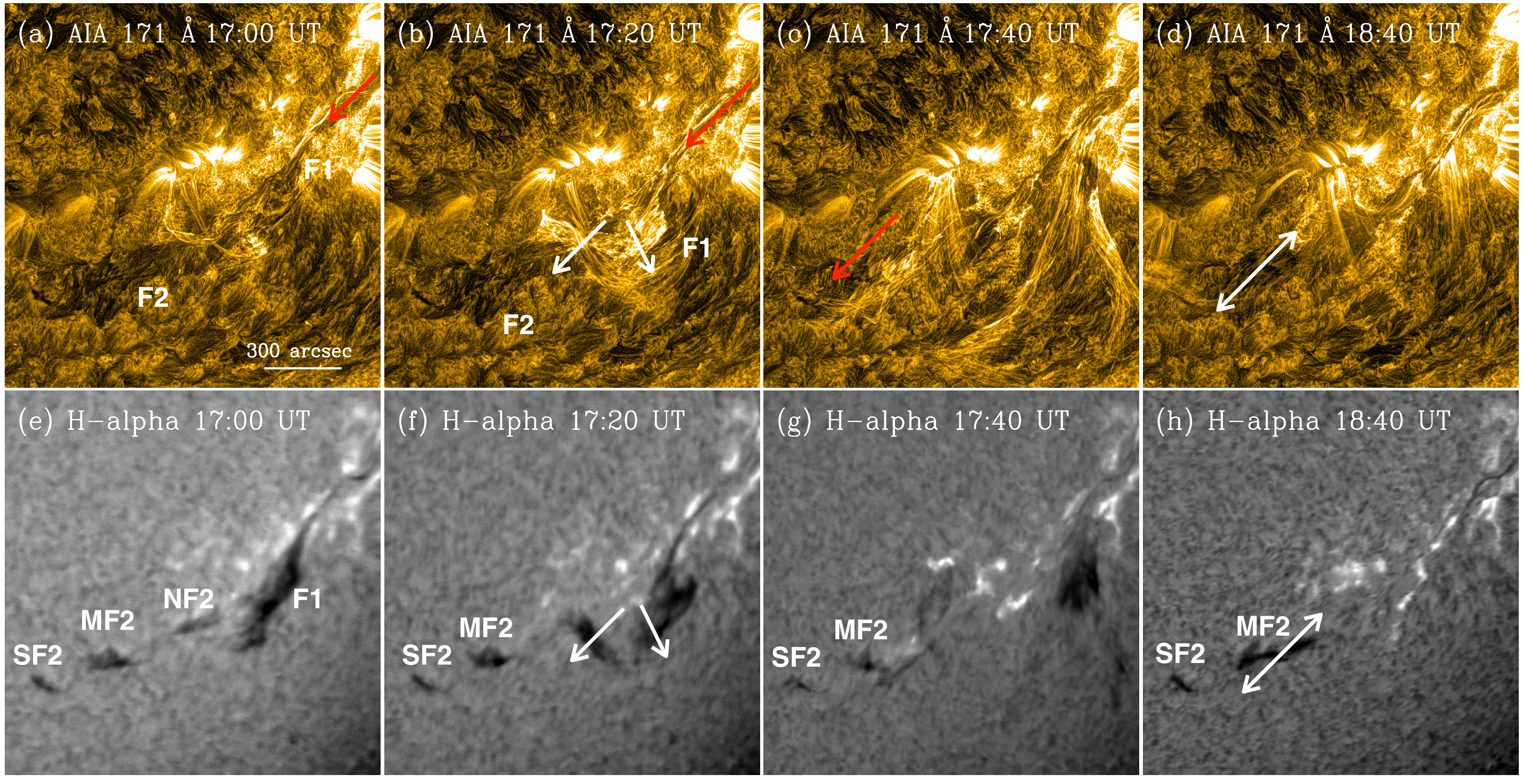}
\caption{Temporal evolution of the filaments observed in SDO/AIA EUV 171 \AA\ (top panels) and GONG H$\alpha$ (bottom panels). For a better visualization of the event we refer to animations associated with Fig. 2 and 3 from \citet{zheng2017}. The locations of the two filaments, F1 and F2, are shown in the panels. In both filters the filaments appear in absorption. In H$\alpha$ the cool plasma is more evident. In panel (e) the three parts of F2 are labeled SF2, MF2, and NF2. The red arrows show the position and direction of the flows seen as bright structures. In (a) the flow emanates from the flaring region, in (b) passes through F1 and in (c) the flow reaches MF2 and SF2. In (b) the white arrow indicates the direction of the motion of the main body of F1. In (f) the white arrows indicate the direction of the motion of the cool plasma. A large portion of the filament moves upwards and finally erupts. The remaining part moves towards the south-eastward direction and interacts with NF2. In (g) the cool plasma from F1 and NF2 reaches MF2, pushing in the direction of the arrow. Finally, in panels (d) and (h) the oscillation is established moving in the direction of the arrow. In this plot, we have used the Multi-Gaussian Normalization (MGN) technique \citep{Morgan14} to make features more clear in the EUV data. This method enhances the solar structures and suppresses the large differences in intensity. 
\label{fig:evolution}}
\end{figure*}

The evolution of the filament eruptions in AIA 171~\AA\ and H$\alpha$ is displayed in Figure \ref{fig:evolution}.
Initially there are two separate filaments, F1 and F2 as named in the study of \cite{zheng2017}. Filament F1 is very large (size $\approx$ 300 Mm) and filament F2 is composed of three differentiated shorter segments (see Fig. \ref{fig:evolution}(a) and (e)). In Figure \ref{fig:evolution}(e) we have labeled the three segments:  SF2, MF2, and NF2 - the South, Middle, and North segments of F2.

Around 17:00 UT filament F1 starts to rise and one part finally erupts. After the rise of F1, a GOES C1.0-class flare is observed. A flare ribbon appears on each side of the polarity inversion line. These two ribbons separate from each other as F1 rises. During the eruption of F1 flows are observed  in 171~\AA\ emanating from the flaring region (Fig. \ref{fig:evolution}(a)) more or less parallel to the spine and passing through F1 (Fig. \ref{fig:evolution}(b)). Around 17:13 UT, the uplifted southern part of F1 comes down and moves south, merging into F2. In Figure \ref{fig:evolution}(f) we see a portion of the cool plasma from F1 approaching NF2 and interacting with NF2, possibly pushing and heating it, as shown in Figure \ref{fig:evolution}(g). Later this flow reaches MF2 and most of the dark structures apparent in EUV disappear (Fig.~\ref{fig:evolution}(c)). In H$\alpha$ MF2 also disappears. These indicate that the plasma from F1 displaces and evaporates almost all the central region of F2. Around 17:50 UT clear brightenings can be seen at western positions in  NF2. These brightenings are probably associated with the hot, erupted plasma reaching the ends of the field lines of the filament channel and heating the chromosphere. At 18:00 UT a mix of bright and dark features moving in opposite directions are observed in 171~\AA. This indicates that part of the filament plasma has been drained but also that part remains in the filament structure. The remaining plasma has reached the maximum displacement and now moves in a northwestern direction, back towards and overshooting its initial location. After 18:15 UT there is no more brightening in the filament as seen in EUV images, indicating that most of the plasma has cooled down. The prominence continues oscillating until approximately 03:00 UT of January 27th. 

The observations suggest that the oscillating plasma in MF2 may be the original plasma pushed by the hot flows coming from F1 and by cool plasma coming from F1 and deposited in MF2. The oscillatory motion is very clear. However, the prominence does not oscillate as a solid body. Initially the motion has some coherence in the filament, but rapidly the motion becomes more complex with counter oscillations in several layers of the prominence. 

The merging of filaments has been reported in previous observations \citep{Chandra11,Filippov16}. Some of the studies also performed MHD simulations of the merging filaments \citep{aulanier2006,Torok11}. In this work, we study the LALOs induced by this merging process, something not reported before.

\section{Analysis of the oscillations}\label{sec:analysis_oscillation}

In this section we study the LAO by the time-distance technique using the H$\alpha$ data from GONG and the 171~\AA\ data from SDO/AIA. In both passbands the filament is seen in absorption, appearing as dark structures against a brighter background. The optical depth of the Lyman continuum near 171~\AA\ and H$\alpha$ are thought to be roughly similar, based on both theory \citep{anzer2005} and observations \citep{schmieder2004a,Heinzel2008a}. However, there are important differences in the appearance of the filaments in the two data sets.
The H$\alpha$ data show the over-all oscillation very clearly because the cool filament plasma appears isolated against the bright disk.  In the 171~\AA\ band images the background is dimmer and more complex, so the filament is not as clear. However, the AIA/SDO instrument offers a better resolution and temporal cadence. In addition, the 171~\AA\ band shows prominence emission thought to be part of the prominence-corona transition region (PCTR) at temperatures $>4\times10^5$~K \citep{Parenti2012}. Like the cool prominence itself, the PCTR is highly filamented, and thus the 171~\AA\ band, with its combination of prominence absorption and emission can highlight fine structure not as obvious in H$\alpha$ or in AIA bands dominated by higher temperature emission, like 193 or 211~\AA.

\subsection{H$\alpha$ data analysis}\label{sec:ha-oscil-analysis}

\begin{figure}[!ht]
\vspace{-2.cm}\centering\includegraphics[width=0.55\textwidth]{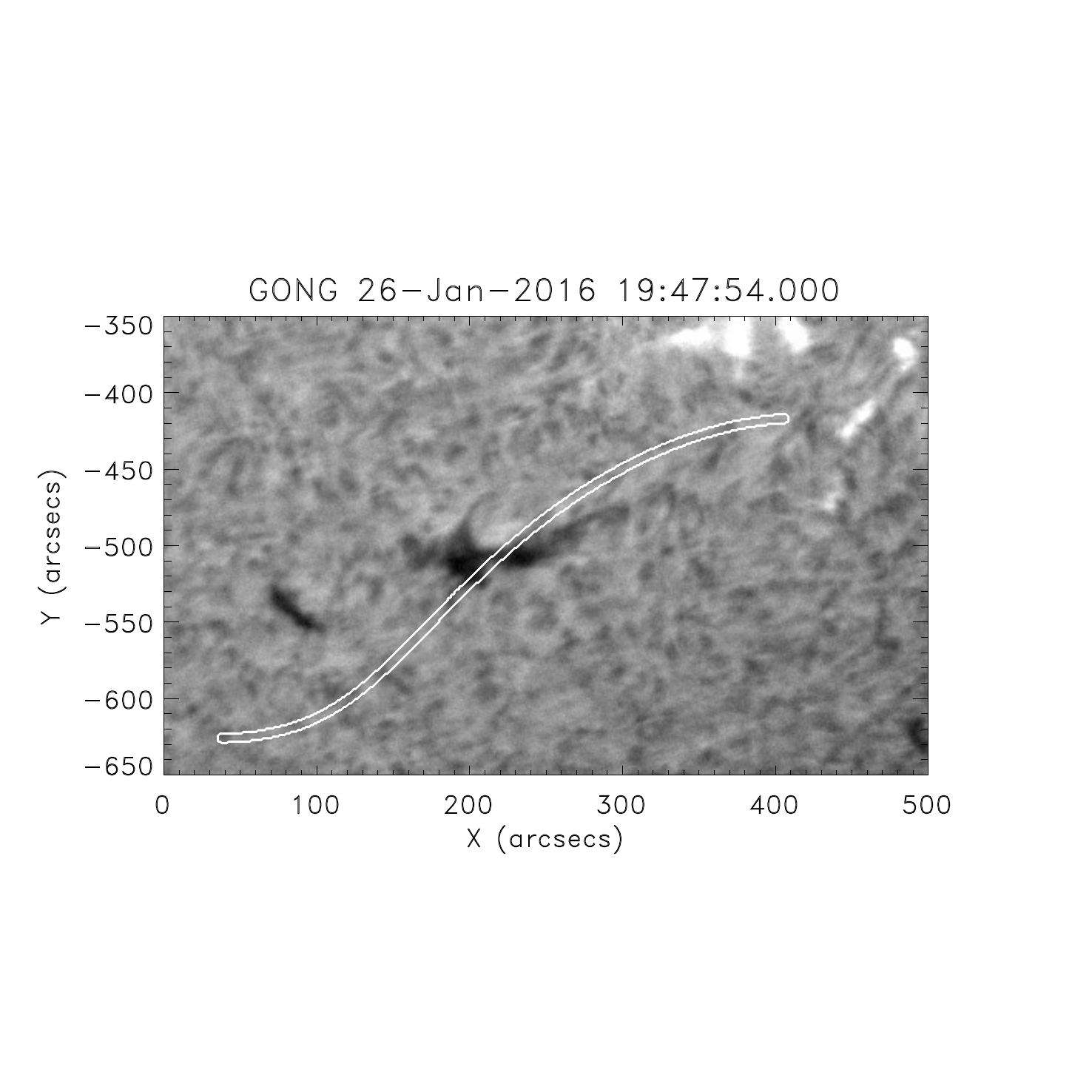}
\vspace{-2.cm}\caption{Plot of the H$\alpha$ data after the eruption of F1. The figure shows the MF2 and SF2 portions of F2. The contour in white shows the artificial curved slit used to construct the time-distance diagram of Figure \ref{fig:gong-data-timedistance}. The path of the slit has been selected by following the motion of MF2 in H$\alpha$. \label{fig:gong-data-map}}
\end{figure}
\begin{figure}[!h]
\centering\includegraphics[width=0.5\textwidth]{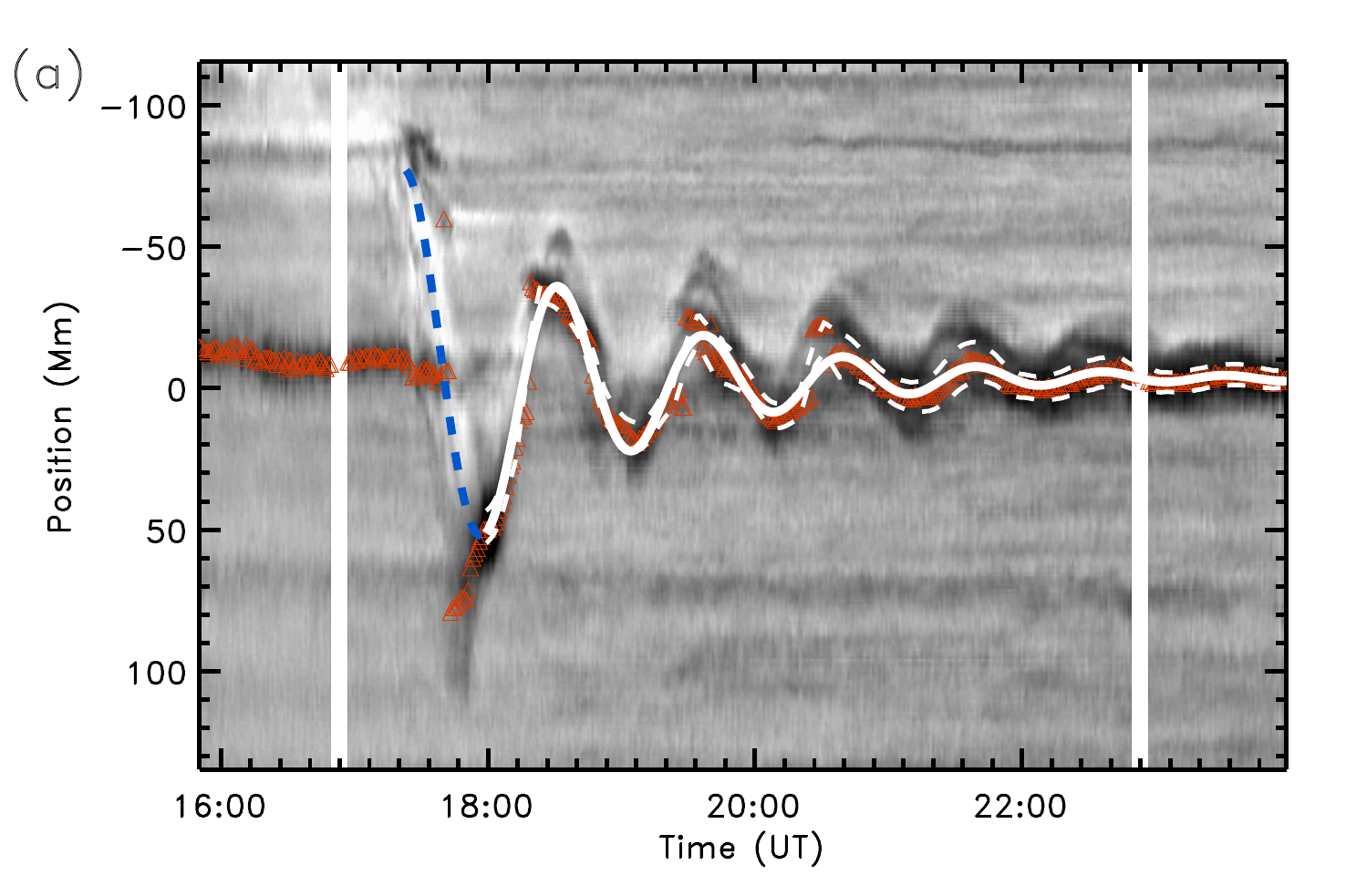}\\
\centering\includegraphics[width=0.5\textwidth]{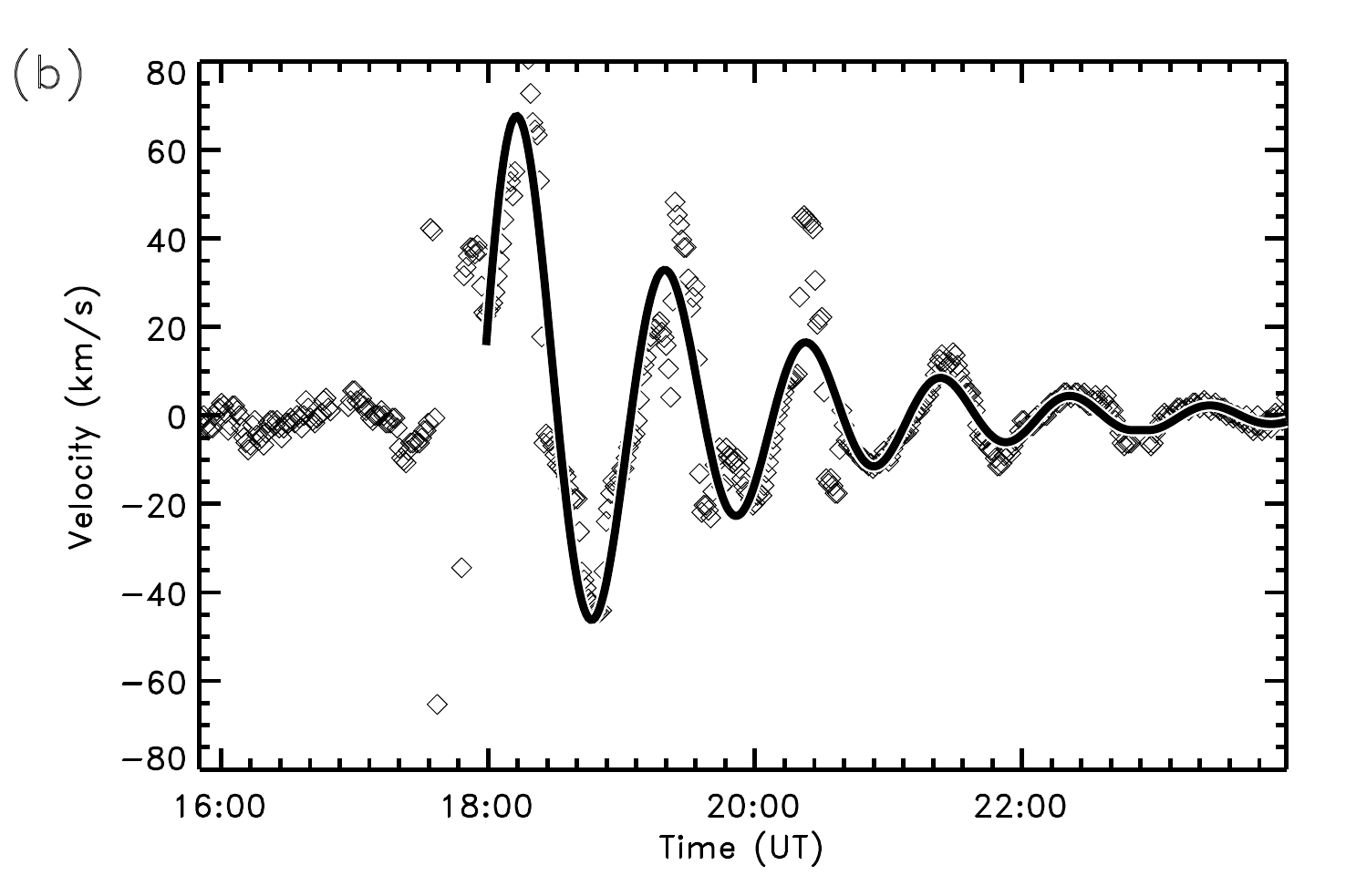}
\caption{(a) Time-distance diagram of H$\alpha$ intensity along the slit shown in Figure \ref{fig:gong-data-map}. The vertical axis represents the position along the slit and the horizontal axis the time in UT. The red dots are $s_0(t)$, the central positions of the dark band (see text). The thick white line is the best fit function of $s_0(t)$ obtained with Equation (\ref{eq:oscillation-fit}). The dashed white lines represent the uncertainty of the position of the filament used to compute the errors of the oscillatory parameters, $\sigma(t)$.  (b) The velocity of the filament computed with the measurements of the position of the dark band in panel (a). The velocity obtained with the fitted function (\ref{eq:oscillation-fit}) is over-plotted as a solid curve. \label{fig:gong-data-timedistance}}
\end{figure}

Figure \ref{fig:gong-data-map}  shows the remaining parts of filament F2 in H$\alpha$ after the eruption of F1.  Over-plotted on the image is the contour of the artificial ``slit'' used to extract the intensity as a function of time along the full temporal range considered. The slit is curved with a width of 6 pixels and length of 402 pixels.  The slit is determined by tracing the motion of the MF2 part of the prominence with the clearest and largest amplitude. The oscillation amplitude is so large that the prominence plasma follows curved trajectories that cannot be followed using straight slits as in our previous work \citep{luna2014}. The resulting time-distance diagram is shown in Figure \ref{fig:gong-data-timedistance}(a). The vertical axis represents the coordinate $s$,  the position along the slit in Mm. The origin of this coordinate is the equilibrium position of the oscillation and the positive direction is along the initial motion of the oscillation, i.e., south-east-ward. 

In the time-distance diagram the grey area is the H$\alpha$ intensity coming from the chromosphere. The white regions are the H$\alpha$ emission from the flaring regions and hot plasma. The dark band corresponds to the position of the cool prominence plasma over the slit. In the positions where the slit crosses the filament the intensity decreases because the prominence absorbs most of the radiation from the chromosphere. In Figure \ref{fig:gong-data-timedistance}(a), before the eruption from 15:45 to 17:25 UT, we see an approximately horizontal dark band associated with an almost steady central segment MF2 at the center of the slit ($s\sim -10$ Mm). We also see NF2 at approximately $s=-80$ Mm. In this time-distance diagram we trace clearly the sequence of the events described in Section \ref{sec:observations}. In the diagram we see that most of the plasma from NF2 is ejected to MF2 after the eruption at 17:25 UT. This plasma is heated and it appears in emission in the H$\alpha$ diagram. It moves from $s=-80$ Mm to $s=0$ Mm reaching MF2 in less than 15 minutes. This ejected plasma interacts with the cool mass of MF2 moving all together to approximately $s=60$ Mm at 18:00 UT. After this time most of the plasma cools down and oscillates around the equilibrium position $s=0$ Mm. 

\begin{figure}
\centering\includegraphics[width=0.5\textwidth]{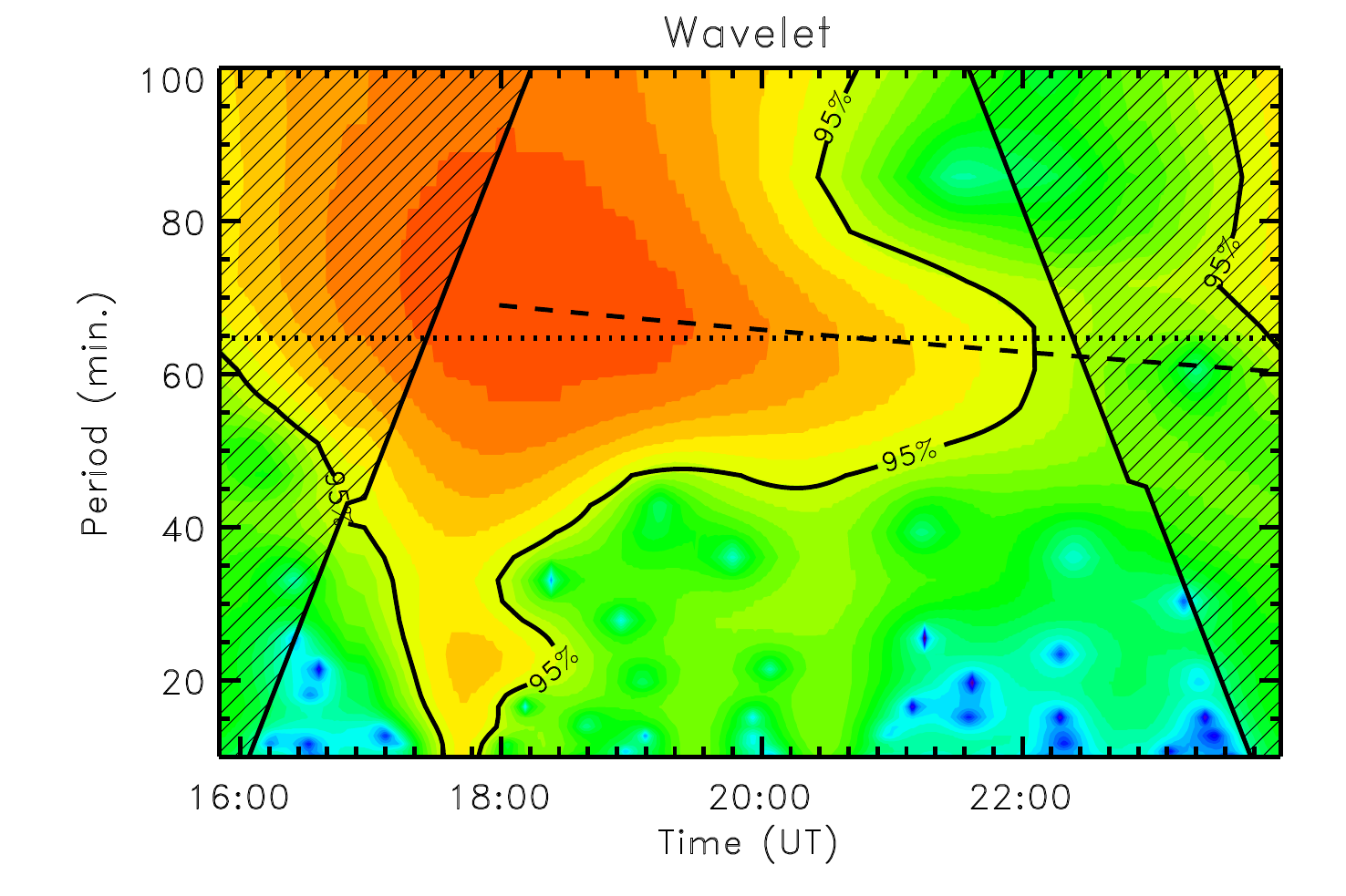}
\caption{Wavelet signal of $s_0(t)$ from H$\alpha$ time-distance. The color code indicates the strength of the wavelet function. The colors from red, orange, yellow, green, blue to dark blue indicate the strength of the wavelet, red being the largest value and dark blue the smallest value. We are not interested in the absolute value of the wavelet and the color scale has been selected for high contrast. The solid thick contour indicates the region where the wavelet function has a value larger than the 95\% of the maximum value. The region encircles the red and yellow region. The shadowed areas are the regions where the wavelet analysis is not valid. The dashed line is the period, $P(t)$ obtained by fitting the oscillation. The horizontal dotted line is the period obtained with the periodogram as discussed in the text.\label{fig:wavelet-ha}}
\end{figure}

In order to study the oscillatory motion of the filament we track the position of the dark band of the time-distance diagram (Fig. \ref{fig:gong-data-timedistance}(a)). We fit the intensity along the slit $I(s,t)$ with a function of the form 
\begin{equation}\label{eq:intensity-fit-function}
I(s,t)=I_0(t) -I_1(t) \, e^{-(s-s_0(t))^2/2\,\sigma(t)^2} \, ,
\end{equation}
for each time $t$. The central position of the dark band is $s_0(t)$ (red dots) and its width is $\sigma(t)$ (white dashed lines). We use the resulting $s_0(t)$ to fit a exponentially decaying sinusoid as
\begin{eqnarray} \nonumber
s_{0, \mathrm{fitted}}(t)=A_0 + A \, e^{-(t-t_0)/\tau}\,\cos&&\left[ \omega(t) (t-t_0)+\varphi_0\right] \\ \label{eq:oscillation-fit}
&&+ d \, (t-t_0) \, ,
\end{eqnarray}
where,
\begin{equation}\label{eq:angular_frequency_linear}
\omega(t)=\frac{2 \pi}{P(t)}=\frac{2 \pi}{P_0}+\alpha \, (t-t_0) \, ,
\end{equation}
$A$ is the amplitude of the oscillation, $t_0$ the initial time of the oscillation, $\tau$ the damping time, $\varphi_0$ the {angular phase at $t_0$} of the oscillation, and $A_0 +d \, (t-t_0)$ is the central position as a function of time. The drift velocity, $d$, accounts for slow displacements not associated with oscillations. In this event the drift velocity is small and the central position is approximately $A_0$. We assume that the angular frequency, $\omega (t)$, depends linearly with time (Eq. (\ref{eq:angular_frequency_linear})). $P_0$ is the initial period at $t=t_0$, and $\alpha$ is the rate of angular frequency variation. The resulting best fit function is plotted in Figure \ref{fig:gong-data-timedistance}(a) as a solid white curve. We find a very good agreement of the fitted function and $s_0(t)$ dots. We assume that all of the uncertainties in the oscillation parameters come from the uncertainty of the determination of the position of the filament. This uncertainty is the width of the Gaussian fit of the H$\alpha$ intensity across the slit, $\sigma (t)$. In Figure \ref{fig:gong-data-timedistance}(a) we have plotted the two curves $s_0(t) \pm \sigma(t)$ that contain the dark band in  H$\alpha$. In the same panel we have also plotted the extrapolated function as a blue dashed line between the flare time at 17:25 UT and $t_0$. In this time range it is not possible to track the motion because there is no clear dark band to follow. However, the extrapolated line corresponds well to  the brightening of the ejected plasma from the northern part of F2. 

In Figure \ref{fig:gong-data-timedistance}(b) the measured velocity is plotted as diamonds. This velocity is computed by the numerical derivative of $s_0$ with time. In general the resulting curve is relatively smooth. However, the measured velocity has several artificial spikes due to jumps in the $s_0(t)$ value. These irregularities occur because the points determining $s_0$ are obtained by manually selecting points on the time-distance diagram and abrupt changes in the first derivative are expected. The spikes are associated with jumps in $s_0(t)$ at approximately 18:15, 19:30, and 20:30 UT.  In this plot we clearly see the large amplitude oscillation with velocities larger than $20\kms$ from the initiation of the oscillation until 21:30 UT. The plot shows very good agreement between the velocity resulting from the fit  (Eq. (\ref{eq:oscillation-fit})) and the measured velocity.

In Table \ref{table:hafit} we show the best fit parameters for Equations (\ref{eq:oscillation-fit}) and (\ref{eq:angular_frequency_linear}). The initial period is $P_0 = 69 \pm 2$ minutes with an angular frequency change rate of $\alpha = 4 \pm 1 \times 10^{-5} \mathrm{\, rad \, min^{-2}}$. This produces a decrease of the period of the oscillation from 69 minutes at 18:00 UT to 60 minutes at 00:00 UT,  corresponding to approximately 1.5 minutes per hour. The Morlet wavelet  \citep{torrence1998} of the position of the filament $s_0(t)$ is plotted in Figure \ref{fig:wavelet-ha}. From the wavelet we see clearly the decrease of the period with time. In the same figure we have also plotted the period obtained with Equation (\ref{eq:angular_frequency_linear}) and values from Table \ref{table:hafit}. The dashed line from the fit follows the period decrease of the wavelet power very well. We have also computed the periodogram \citep{lomb1976} using the algorithm by \citet{carbonell1991}. The method is equivalent to a Fast-Fourier transform analysis. We obtain a value of $65 \pm 2$ minutes,  an intermediate value between the maximum and minimum period $P(t)$. The periods obtained are larger than the 55 minutes obtained by \citet{zheng2017}. The change in the period could be due to changes in the local prominence field geometry as suggested by \citet{bi2014} or an increase of the coronal temperature, as suggested by \citet{ballester2016}. 

\begin{table}[h!]
\centering
 \begin{tabular}{l c}
 \hline
Period, $P_0$ (min.) & 69 $\pm$ 2 \\
Frequency drift, $\alpha$ (rad./min$^2$) & 4 $\pm$ 1 $\times 10^{-5}$ \\
Period periodogram (min.) & 65 $\pm$ 2 \\
Damping Time, $\tau$ (min.) &  85 $\pm$ 7\\
Amplitude, $A$ (Mm) & 52 $\pm$ 3 \\ 
Velocity Amplitude ($\kms$) & 68 $\pm$ 3 \\
Max. Vel. extrapolated ($\kms$) & 100 $\pm$ 3 \\ [0.1ex] 
 \hline
 \end{tabular}
 \caption{Best fit parameters of H$\alpha$ time-distance diagram. The maximum velocity extrapolated refers to the velocity obtained from the blue dashed line of Fig. \ref{fig:gong-data-timedistance}(a).}
 \label{table:hafit}
\end{table}
We find a velocity amplitude of $68 \pm 3 \kms$ that is larger than the value obtained by  \citet{zheng2017}. This is because our method follows the motion of the cool plasma along curved paths in contrast with the straight slits used in the previous study. The maximum velocity of the extrapolated function is $100 \pm 3 \kms$. This velocity is associated with the motion of the flows ejected from F1. This value is clearly below the $165 \kms$ obtained by  Zheng et al.

The damping time is $\tau=85 \pm 7$ minutes, similar to the period and thus indicating a strong damping of the motion. The strength of the damping is measured by the ratio $\tau/P = 1.2 \pm 0.1$ where we have used the maximum period $P_0=69 \pm 2$ minutes. This value indicates that the oscillation reduces a factor $e$ in almost one period. Several damping mechanisms have been proposed to explain such strong damping \citep[see, e.g.,][]{luna2012b,Zhang2013a,ruderman2016}.

\subsection{171\AA\ data analysis}\label{sec:171-oscil-analysis}

\begin{figure}[!ht]
\centering\includegraphics[width=0.5\textwidth]{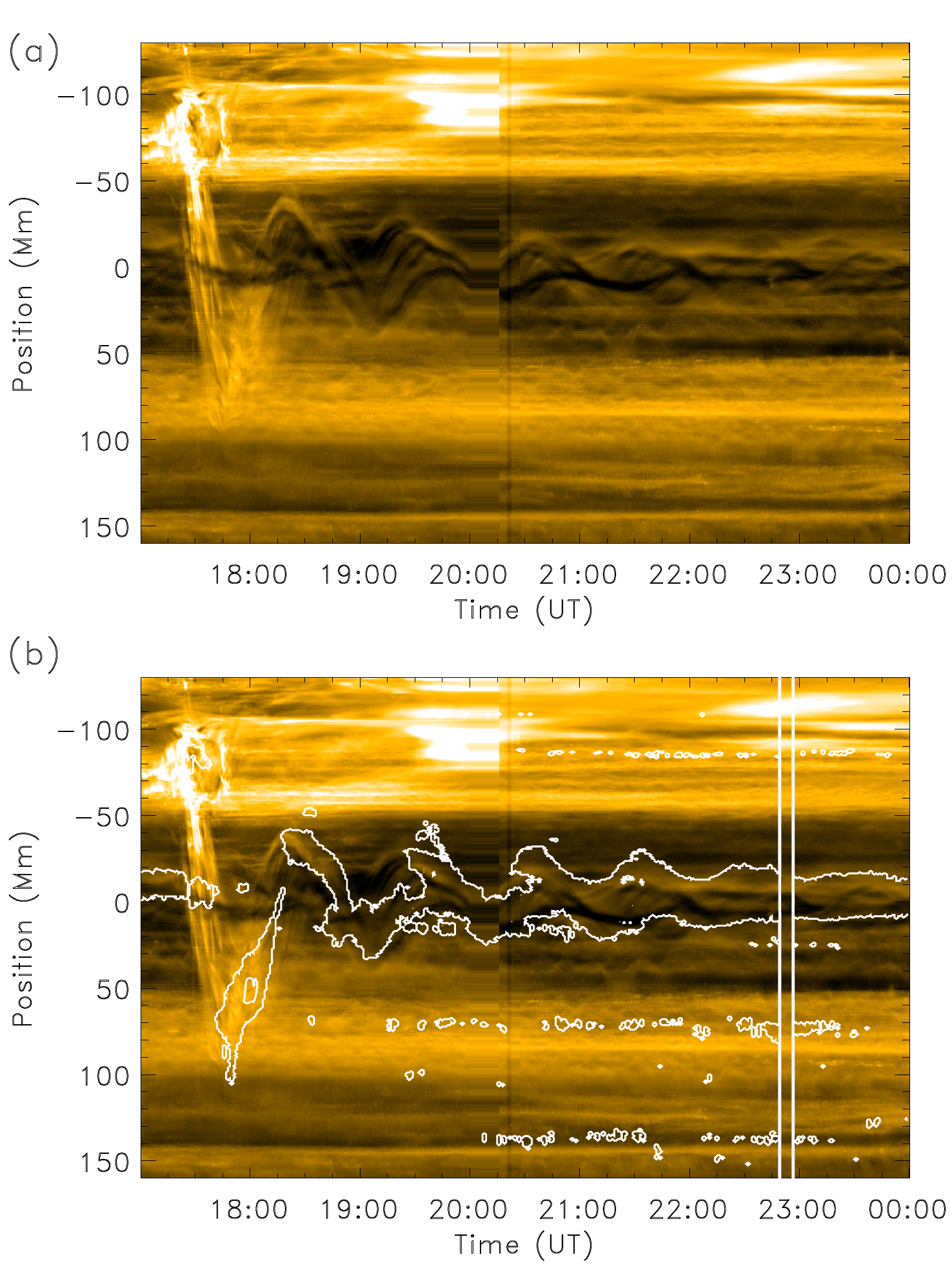}
\caption{(a) Time-distance diagram in 171\AA\ filter in panel. (b) The same diagram over-plotted with contours from the H$\alpha$ time distance diagram from Fig.~\ref{fig:gong-data-timedistance}. There is a large data gap between 20:00 and 20:15 UT. We have used the intensity at 20:00~UT to fill this gap.  \label{fig:likeHa-timedistance}}
\end{figure}

The 171~\AA\ data offer more details of the motion than the H$\alpha$ data due the combination of better resolution, better time cadence and emission from the PCTR (see Sec~\ref{sec:analysis_oscillation}).  In order to understand the motion of the {warm plasma emitting in} 171\AA\ with respect to the cool plasma in H$\alpha$ data we first generate the time-distance diagram in 171\AA\ using the same slit used for the H$\alpha$ analysis (Fig. \ref{fig:gong-data-map}). Figure \ref{fig:likeHa-timedistance}(a) shows the resulting time-distance diagram in 171\AA, while  Figure \ref{fig:likeHa-timedistance}(b) shows the same image over-plotted the H$\alpha$ intensity contour.  

At the {triggering} phase of the oscillation the interaction of the hot flows from F1 with NF2 is seen as a bright spot at 17:25 UT. After the interaction we can see the hot plasma in emission flowing along the slit. A more complete analysis of the {triggering} phase is given below. After the {triggering} phase, $t >$ 18:00 UT,  several thin dark bands appear inside the oscillating structure (see Fig. \ref{fig:likeHa-timedistance}(a)). Each dark band is probably associated with an individual thread or small bundle of threads. In addition, at the top and the bottom of each dark band faint emission can be seen associated with the PCTR between the cool thread and the hot corona. Each dark band (i.e., prominence thread) shows an identical pattern of absorption and PCTR. Using this pattern of absorption and faint emission, we can see several threads oscillating around the equilibrium position at $s\sim0$ with different phases and amplitudes. The threads are aligned in more or less the same vertical direction. The combination of the complexity of the motion with the absorption-emission pattern makes it impossible to track the motion of an individual thread. From the figure we see that only half a period can be followed in the diagram.

In Figure \ref{fig:likeHa-timedistance}(b) we can see the correspondence between the H$\alpha$ and 171~\AA\ data sets. The H$\alpha$ dark-band does not correspond with any of the thinner individual threads observed in 171~\AA, but instead envelops all the 171\AA\ threads. This indicates that the H$\alpha$ dark-band is the average motion of several threads oscillating out of phase. Before 20:00 UT the dark-bands are more or less parallel in the diagram, indicating that thread velocities are similar. However, the periods are different because they reach the maximum elongation at different times (at $\sim$ 19:30 UT). After 20:00 UT the phase differences are more evident. In fact, around 22:00 UT threads oscillating with opposite velocities are seen and there is less correspondence between the two filters. Although the oscillation envelope apparent in the H$\alpha$ time-distance diagram is visible in 171~\AA, it are not as apparent, perhaps because of spatially varying foreground emission. The combination in 171~\AA\ of emission and absorption shows a more complex, multi-threaded structure than the H$\alpha$. This indicates that even though the plasma seen at the different wavelengths is moving similarly, different structures are emphasized. The lower spatial resolution of the H$\alpha$ data also contributes to the discrepancies because the lower resolution reduces visibility of fine structures.

From this figure we see that in the H$\alpha$ time-distance diagram the dark band represents in some sense the bulk or averaged oscillation of the threads with different periods. \citet{luna2016a} predicted such phase differences in different layers of the prominence. As we see in Figure \ref{fig:likeHa-timedistance}(a) the different threads reach their maximum elongation at different times,  broadening the position of the maximum. This is very clear between the maximum at 18:20 UT and the one at approximately 19:30 UT. In the first the maximum is very clear because the threads reach the maximum at more or less the same time. However, due to phase differences the threads reach the second maximum at different times. This effect could reduce the period measured from the H$\alpha$ time-distance diagram.

\begin{figure}
\vspace{-0.cm}\includegraphics[width=0.5\textwidth]{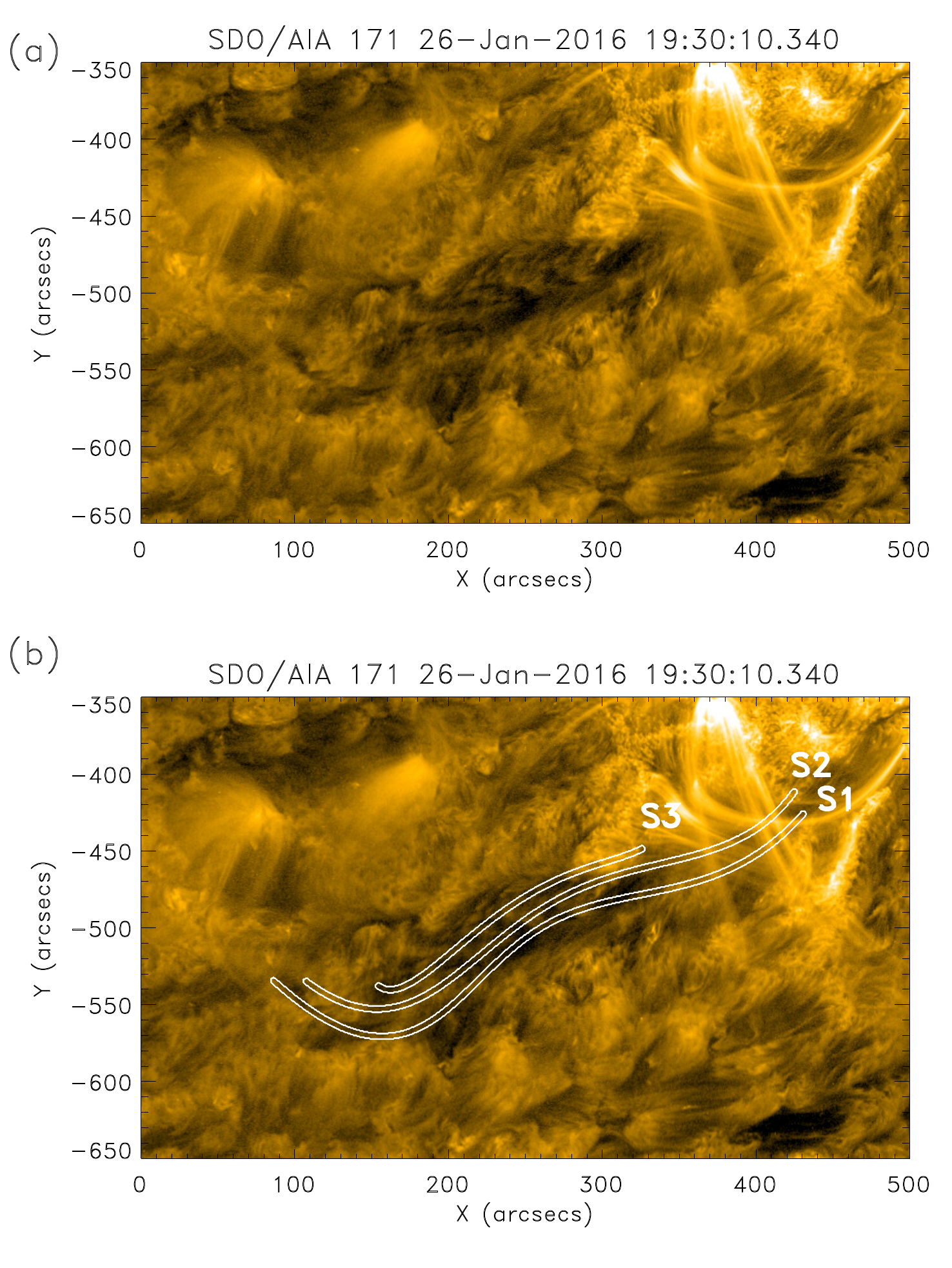}
\caption{(a) The filament channel in the 171~\AA\ band. At the center of the image the central portion of the F2 filament appears in absorption. (b) the same image over-plotted with the three artificial slits used to study the oscillatory motion. \label{fig:euv171mapwithslits}}\vspace{0.4cm}
\end{figure}

Now we extend our analysis of the 171~\AA\ data to several slits placed at different positions in the filament. As discussed above, from 171~\AA\ images we can perceive more details of the motion of the prominence than from H$\alpha$ data. As in the H$\alpha$ analysis, we track the motion of clear features to define the artificial slits. We have defined 3 slits over the main body of the filament. In Figure \ref{fig:euv171mapwithslits} we have plotted one frame of the AIA 171~\AA\ and the 3 artificial slits used, labeled as S1 to S3. They have been defined by tracking by eye the motion of distinguishable features in curved paths. The width of each slit is 4 pixels. Our analysis is focused on the central part of the filament channel shown in Figure \ref{fig:euv171mapwithslits}. The temporal sequence analyzed starts at 17:00 UT January 26 and ends at 05:00 UT January 27.

\begin{figure}[!h]
\vspace{-0.cm}\includegraphics[width=0.4\textwidth]{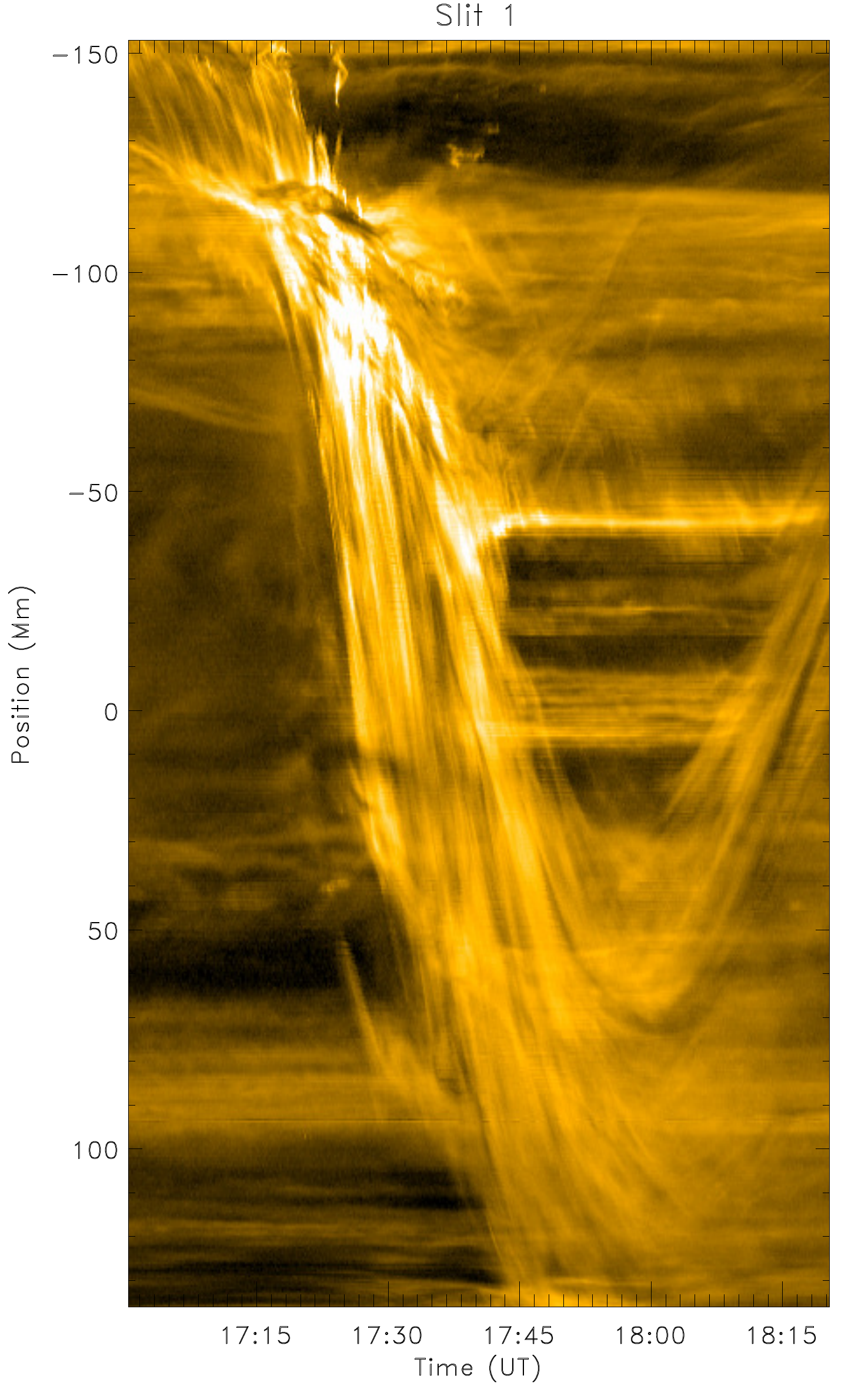}
\caption{Time-distance diagram of the {triggering} phase in S1.\label{fig:initialphase171}}\vspace{0.4cm}
\end{figure}

In Figure \ref{fig:initialphase171} we have plotted the time-distance diagram of the {triggering} stage in S1. In this figure we can see the events described in Section \ref{sec:observations}.  Hot flows coming from the flaring region and F1 are seen as bright features before 17:20 UT and $s>120$ Mm (see also red arrow in Fig. \ref{fig:evolution}(b)). These flows interact with NF2, producing a bright spot in the diagram between 17:20 and 17:30 UT and between -120 and -80 along the slit. Dark structures can also be seen emanating from NF2. This indicates that the flows from northern positions heat up and push the cool plasma of NF2. After 17:30~UT a mix of bright and dark features can be identified. This flow reaches the central segment of F2, pushing the cool plasma located at approximately $s \sim 10$ Mm (in correspondence with Fig. \ref{fig:evolution}(c)). This initial cool plasma is very faint in Figure \ref{fig:initialphase171}. It is more clear in Figure \ref{fig:gong-data-timedistance}(a). There is not a clear increase of the brightening at this point, indicating that the flows from northern positions only push the cool plasma of MF2 and do not heat it.  The intensity of the bright structures decreases with time and distance relative to the bright spot indicating that the hot flows are cooling down. After 18:00 UT most of the plasma has cooled down and we can see the plasma in absorption in 171~\AA\ and in H$\alpha$ (Fig. \ref{fig:gong-data-timedistance}(a)). Also, at 18:00 UT the plasma reaches a maximum elongation and after this time the plasma moves backwards, establishing the oscillation.  

\begin{figure}[!ht]
\includegraphics[width=0.5\textwidth]{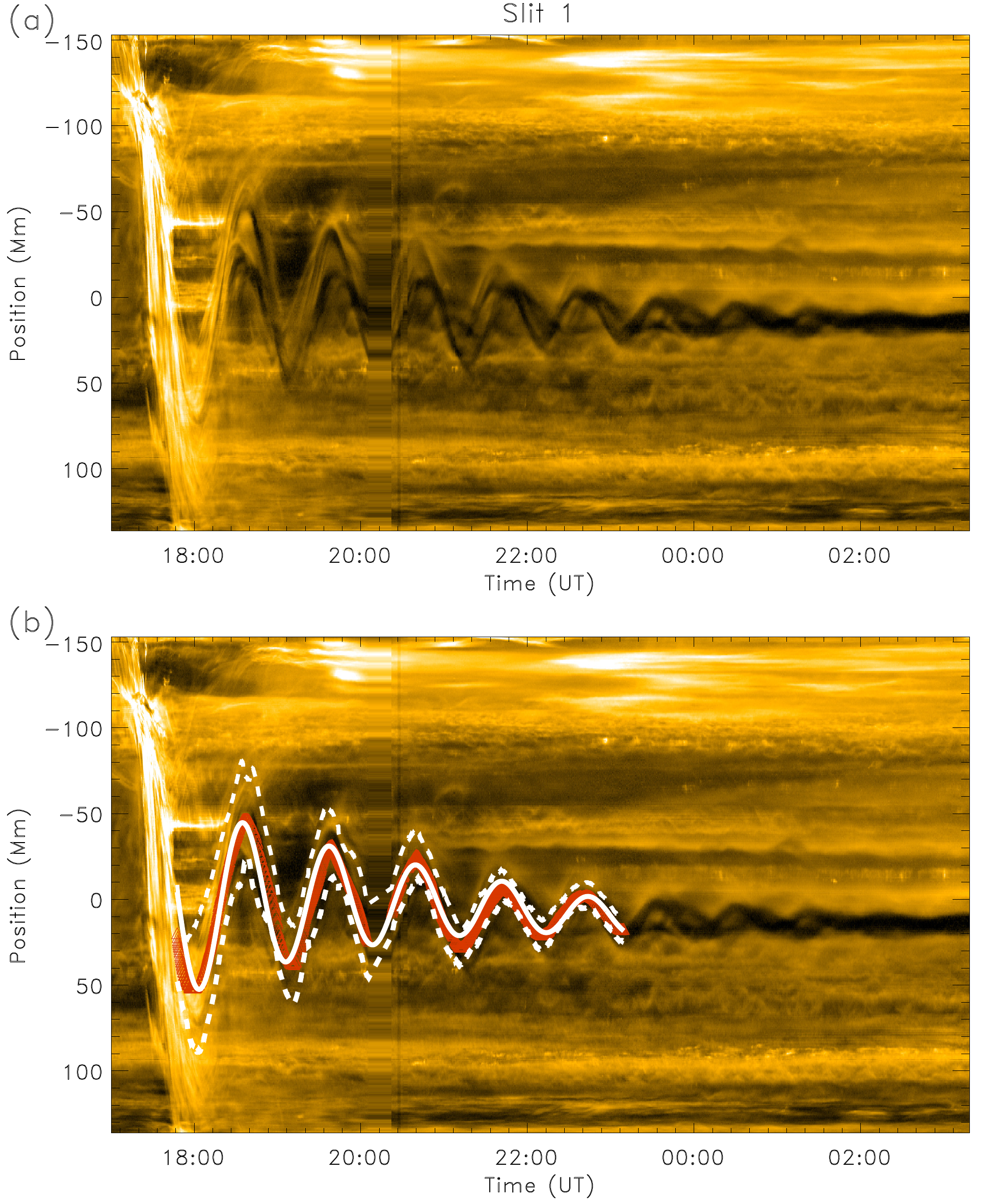}
\caption{(a) Time-distance diagram from S1. (b) The same diagram with the best fit oscillation over-plotted as a solid white line. The red line shows the $s_0$ positions. The dashed white lines indicate the region of uncertainty given by $s_0\pm\sigma$. \label{fig:timedistance_EUV_slit1}}
\end{figure}

{Figure \ref{fig:timedistance_EUV_slit1} shows the time-distance diagram for S1. As we have discussed, hot flows reach the filament and trigger its oscillations. The oscillatory pattern is very clear in the time-distance diagram of S1. The oscillation starts at 17:25~UT and ends around 02:00 UT, lasting for more than 8 hours. In the time-distance diagrams of S2 and S3 the oscillatory motion is also clear in a similar temporal range. However, the motion is more complex with many threads oscillating with different periods and phases. {The time-distance diagrams for S2 and S3 are not shown because they do not give additional information. However, the results of the fits are given} in Table \ref{table:aiafit}.}

For the H$\alpha$ data we have used the minimum of the intensity to track the motion of the prominence and study the oscillatory parameters. Due to the complexity of the motion of this event it is impossible to follow the motion of only one thread in 171\AA. In the first stage of the oscillation, lots of bright threads are observed with few thin dark bands that appear and disappear (see Fig. \ref{fig:initialphase171}). After the {triggering} phase clear dark threads are seen. In the time-distance diagram several threads cross each other and it is not possible to identify the motion of just one thread and follow the motion of one continuous dark-band as in the case of the H$\alpha$ data. 

This difficulty forces us to change the method of tracing the oscillation. Although the details are complex, a global oscillation can still be distinguished in the time-distance diagrams of Figure \ref{fig:timedistance_EUV_slit1} {and the diagrams corresponding to S2 and S3}. {It consists} of a bundle of bright and dark threads at the beginning of the motion or a bundle of dark threads with emission in the PCTR later on. To trace this we first define a curve {that defines the central position of the bundle of threads by selecting points along the time-distance diagrams}. The uncertainty is defined as the total width of the bundle of light and dark oscillating threads. We then fit these values using Equation (\ref{eq:oscillation-fit}) to obtain a curve defining the central position of the oscillation.

 In Figure \ref{fig:timedistance_EUV_slit1}(b) we have plotted the selected $s_0$  positions as a red line. The uncertainty limits, $\sigma$, are plotted as white dashed lines. Finally, the best fit oscillation curve is plotted as a solid white line. In the fit of oscillations we have assumed that the period of the oscillation is constant with $\alpha=0$ (see Equation (\ref{eq:angular_frequency_linear})). This is because there is not a clear constant variation of the period with time. {The fitted values (red curve) do not cover the entire temporal domain of each slit. In S1 we have fitted $s_0$ up to approximately 23:00 UT. The oscillation is clear up to the end of the temporal domain, but the fitting fails when including points beyond 23:00 UT. For the other time-distance diagrams the oscillation is only clear in a part of the temporal domain. For these cases we have fitted the $s_0$ when the oscillation is clear.}

The fitted function represents  the global motion of the prominence very well. In Table \ref{table:aiafit} the best fit parameters and corresponding errors are summarized.
\begin{table}[h!]
\centering
 \begin{tabular}{c | c | c | c | c} 
 & $P$ (min) & $\tau$ (min) & $A$ (Mm) & $V$ ($\kms$)\\ [0.5ex] 
 \hline
Slit 1 & 60.5 $\pm$ 0.2 & 170 $\pm$ 11 & 58 $\pm$ 3  & 100 $\pm$ 7\\
\hline
Slit 2 & 64.3 $\pm$ 0.4 & 209 $\pm$ 22 & 60 $\pm$ 4 & 95 $\pm$ 5\\
\hline
Slit 3 & 68.5 $\pm$ 0.4 & 254 $\pm$ 34 & 26 $\pm$ 2 & 40 $\pm$ 5 \\ [0.ex] 
 \hline
 \end{tabular}
 \caption{Best fit parameters from AIA/171\AA\ time-distance diagrams in slits 1 - 3.}
 \label{table:aiafit}
\end{table}
The periods have values around one hour. The damping time in S1 is $170$ minutes indicating a damping time per period ratio of $\tau/P=2.8 \pm 0.1$. This value is over twice that derived from the H$\alpha$ data. {For S2 and S3 the damping time is even larger, with $\tau/P=$3.2 $\pm$ 0.1, and 3.7 $\pm$ 0.2 respectively}. The amplitude of the oscillation in S1 and S2 is around 60 Mm. The total displacement is 116 Mm, a non-negligible portion of a solar radius.  In S1 the velocity amplitude is 100 $\pm$ 7 $\kms$. This the largest value reported thus far for a LAO, being slightly larger than the previous highest value of 92 $\kms$, reported by \citet{jing2003}. For S2 the velocity amplitude is also considerable with 95 $\pm$ 5 $\kms$. {In contrast, for S3 both the amplitudes of the displacement and velocity are considerably smaller.} 

\begin{figure}[!h]
\includegraphics[width=0.5\textwidth]{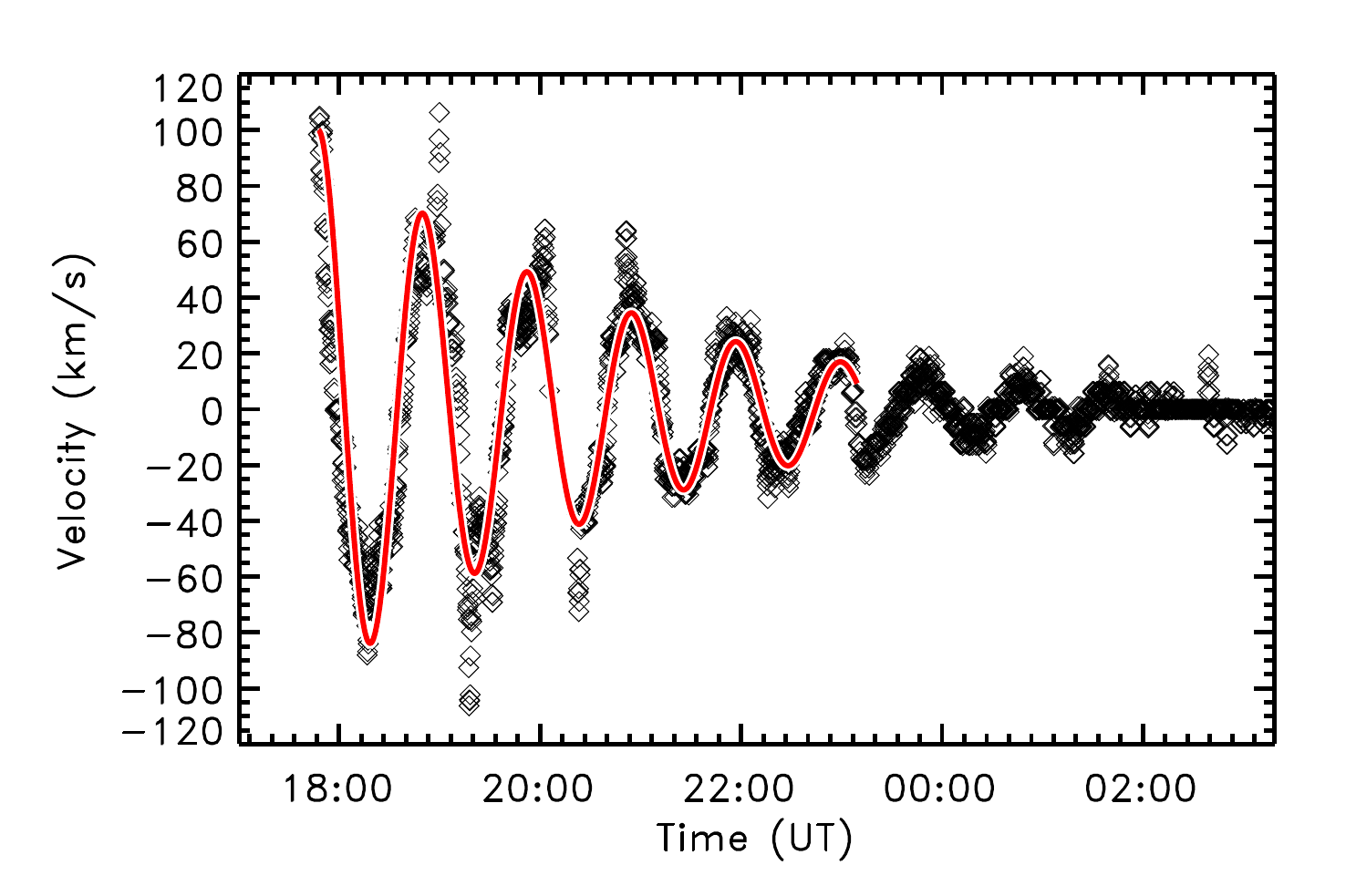}
\caption{The same as Figure \ref{fig:gong-data-timedistance}(b) but for the data taken from the AIA 171~\AA\ time-distance diagram for S1. The data shown starts at the same initial time as the fitted data in Figure \ref{fig:timedistance_EUV_slit1}(b) and ends at the end of the full temporal sequence. The red solid line is the velocity obtained using the fit to the oscillation (Eq. (\ref{eq:oscillation-fit})). It corresponds to the temporal derivative of the white solid line of Fig. \ref{fig:timedistance_EUV_slit1}(a).}\label{fig:AIA_velocity_measured}
\end{figure}

We compute the velocity using $s_0(t)$ data from the S1 time-distance diagram and calculating the numerical temporal derivative as in the H$\alpha$ data analysis. In Figure \ref{fig:AIA_velocity_measured} the resulting velocity has been plotted as diamonds. As with the analysis of the H$\alpha$ data, in the plot there are 5 spikes associated with irregularities in the $s_0(t)$ data and that do not reflect real velocities, as is discussed in Section~\ref{sec:ha-oscil-analysis}. These spikes in the velocity are at approximately 18:50, 17:20, 20:00, 20:33, and 21:00 UT. In this plot we can see the large velocities involved in this event. The velocity measured from the time-distance of the oscillation remains larger than $20\kms$ for 6 hours after the triggering up to 00:00 UT. In the figure we also show the velocity resulting from the fit (Eq. (\ref{eq:oscillation-fit})) with a red solid line. The velocity derived from the oscillation fit matches the measured velocity very well.

\begin{figure}[!h]
\centering\includegraphics[width=0.5\textwidth]{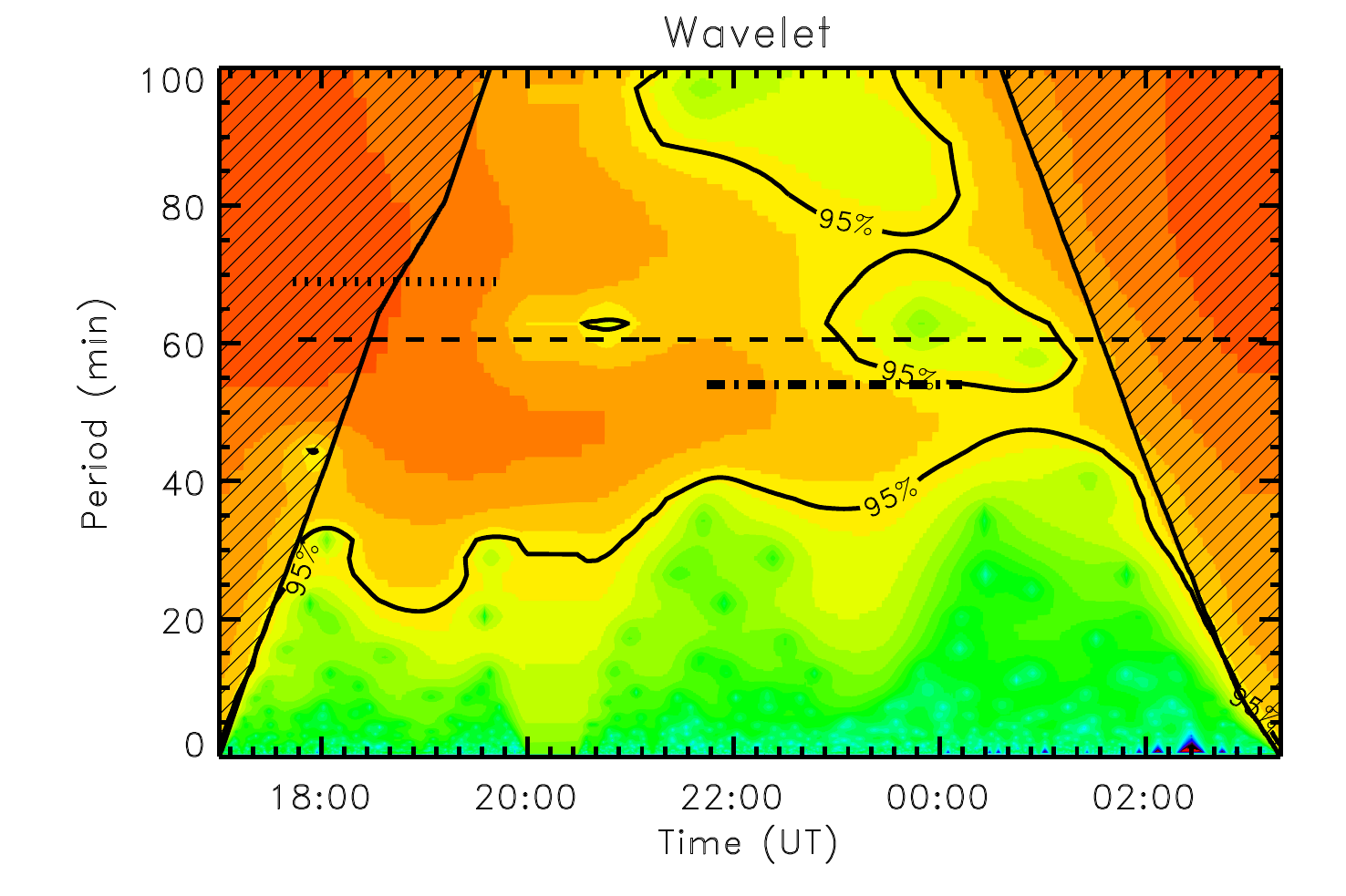}
\caption{As in Figure \ref{fig:wavelet-ha}, the wavelet signal for $s_0(t)$ from S1. The period obtained from the oscillation fit is plotted as a dashed horizontal line. The periods of the fits of the initial and final phases from Fig. \ref{fig:twophases-fit-s1} are plotted as dotted and dot-dashed horizontal lines respectively. These lines are also plotted in the temporal domain of the respective phases. The large period corresponds to the fit associated to the initial stage of the oscillation while the smaller one corresponds to the period of the final stage of the oscillation. \label{fig:wavelet_S1}}
\end{figure}

Using the $s_0(t)$ positions from the S1 we can analyze the evolution with time of the period of the oscillation using the wavelet analysis as we have done for the H$\alpha$ data. We have focused on the S1 time-distance diagram because the oscillation is very clear for a large number of periods after the triggering. Figure \ref{fig:wavelet_S1} shows the wavelet power as a function of time. We clearly see that at the beginning of the oscillation the wavelet power has a maximum signal centered at approximately 70 minutes (red). However, after 18:30 UT there is a second period close to 50 minutes that remains more or less constant up to the end of the temporal series. {Thus, the oscillation period is long only at the beginning of the event but rapidly declines later.} This behavior contrasts with the H$\alpha$ data where the period decreases monotonically with time along the full temporal sequence.  

\begin{figure}[!h]
\includegraphics[width=0.5\textwidth]{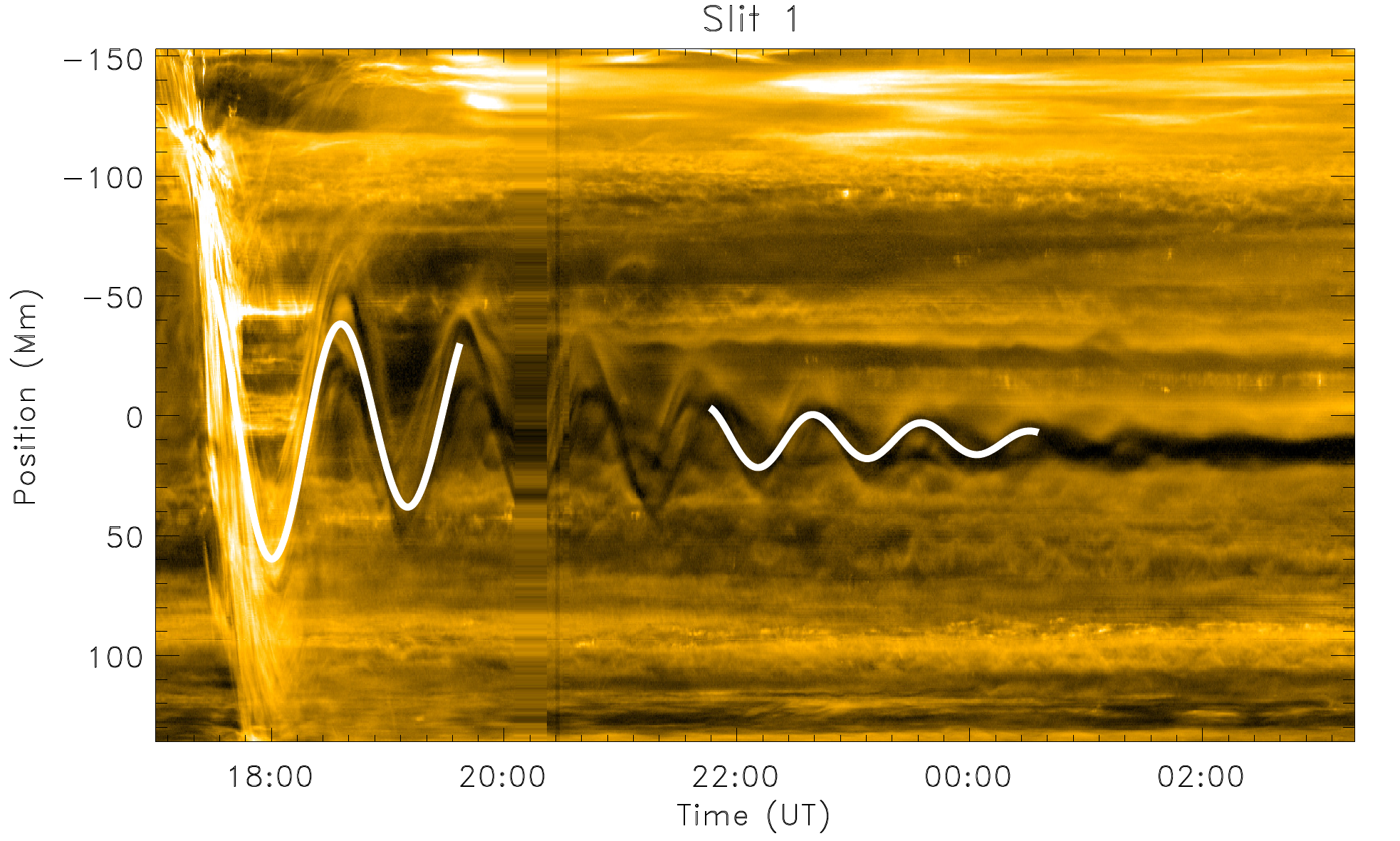}
\caption{Time-distance diagram of S1 and the best fit functions of the initial and final phases of the oscillations (white solid curves).\label{fig:twophases-fit-s1}}
\end{figure}

In order to parameterize this behavior we have fitted a function of the form Equation (\ref{eq:oscillation-fit}) with $\alpha=0$ for {an} initial phase of the event (17:30 UT to about 19:30 UT) and in a later phase (21:45 to 00:40 UT). Figure \ref{fig:twophases-fit-s1} shows the two fitted functions in the two phases of the oscillation, and Table \ref{table:twophasesfit} summarizes the results of the fits. The period of the first phase is $69 \pm 2$ minutes, in agreement with the initial period found for H$\alpha$ oscillation. In the second phase the period is considerably smaller, $54 \pm 1$ minutes. The two periods are over-plotted in Figure \ref{fig:wavelet_S1} as horizontal dotted lines. The result of the analysis of the two phases agrees with the wavelet analysis. The first phase period is more or less the averaged period of the wavelet signal in the time range of the initial phase. The last phase period coincides also with the average value of the wavelet signal in the temporal domain of the final phase of the oscillation.

Thus, this oscillation is characterized by an initial stage with a long period and a posterior oscillation with a smaller period. This behavior can be associated with the flare and eruption that triggers the oscillation. The flare and eruption produce hot flows of plasma with high pressure on the side of F1. This pressure imbalance between both sides of the prominence pushes the cool threads from the equilibrium position. The high pressure associated with the flaring region is not impulsive in the sense that the duration is comparable to the period of the oscillation. This means that in the first phase of the oscillation there is no free oscillation of the system. The high pressure gradients act in the opposite direction of the gravity projected along the field producing an increase of the period in the first half of the oscillation cycle. {A second explanation is that the magnetic field evolves during the first stage of the oscillation. After the eruption and flare the magnetic field may still be relaxing to a more stable configuration. An alternative explanation for the change of field geometry is that part of the heavy prominence mass from F1 and NF2 is transferred to MF2. Due the increase of the prominence mass in MF2, the depth of the dips that support the prominence mass may also increase. In both cases the change of the field geometry would be reflected in the oscillations.}

\begin{table}[h!]
\centering
{\small
 \begin{tabular}{c | c | c | c | c} 
 & $P$ (min) & $\tau$ (min) & $A$ (Mm) & $V$ ($\kms$)\\ [0.5ex] 
 \hline
Initial phase & 69 $\pm$ 2 & 200 $\pm$ 60 & 62 $\pm$ 5  & 88 $\pm$ 9\\
\hline
Final phase & 54 $\pm$ 1 & 151 $\pm$ 30 & 14 $\pm$ 2 & 25 $\pm$ 4\\ [1ex] 
 \hline
 \end{tabular}}
 \caption{Best fit parameters resulting from the fits in the two stages of the oscillation using 171~\AA\ data.}
 \label{table:twophasesfit}
\end{table}

The damping times, $\tau$, at both phases of the oscillation are consistent within uncertainties. These values are also in agreement with the values in Table \ref{table:aiafit}. The amplitudes and velocity amplitudes are also consistent with the previous fits and with the fact that the motion is damped and the amplitude of the motion reduced.

The damping times obtained in the analysis of the EUV S1 time-distance diagram are roughly twice the values obtained in the analysis of H$\alpha$. This indicates that the oscillation analyzed in H$\alpha$ shows stronger damping than the same events analyzed in 171~\AA. The reason for this discrepancy could be that in H$\alpha$ time-distance diagrams we are observing bulk or averaged motion of several threads moving with different phases. The loss in coherence of the threads produces an apparent increase in damping measured from the H$\alpha$ data. 

\section{Magnetic field modeling} \label{sec:flux-rope-model}

To understand the magnetic structure supporting the erupting filament, we construct magnetic field models using the flux-rope insertion method developed by van Ballegooijen \citep{vanballegooijen2004}. We briefly introduce the method below; for detailed descriptions refer to \citet{bobra2008} and \citet{su2009,su2011}.

\begin{figure}[ht!]
\centering\includegraphics[width=0.47\textwidth]{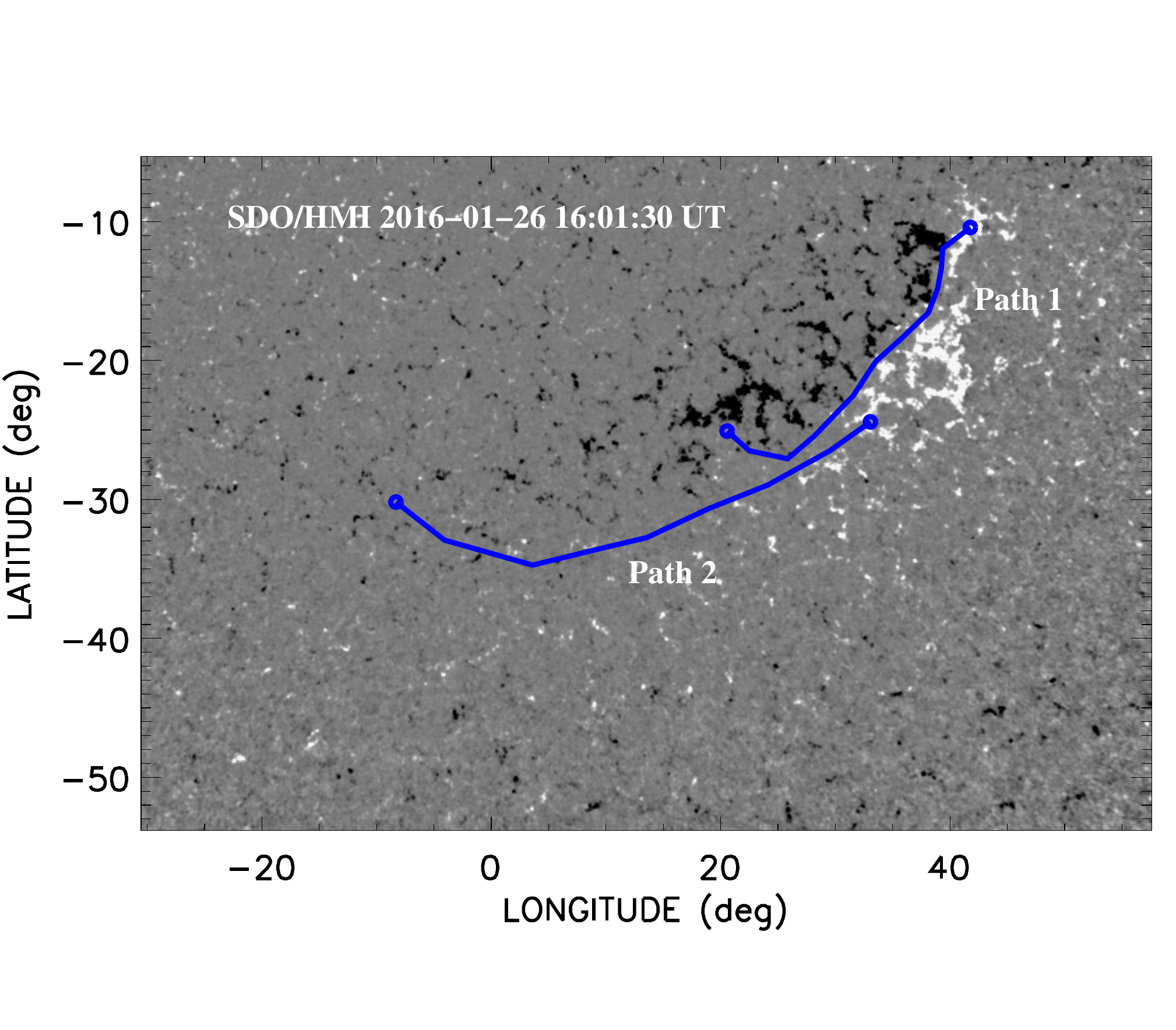}
\caption{The longitude--latitude map of the radial component of the photospheric magnetic field by SDO/HMI in the HIRES region at 16:01 UT on 2016 January 26. The blue curves with circles at the two ends show the paths along which we insert flux ropes.
\label{model-path}}
\end{figure}

At first, a potential field model (HIRES) is computed from the high-resolution  and global magnetic maps observed with SDO/HMI. The lower boundary condition for the HIRES region is derived from the photospheric line-of-sight magnetograms obtained at 16:01 UT on 2016 January 26. This time is slightly before the events we analyze, at the time when the region was at the center of the disk. We do not expect the magnetic field to have changed greatly over the next half day, and, in fact, similar filament analyses have used magnetograms from a week before or after to model magnetic structures at the limb \citep{su2012b,jibben2016,su2015}. The longitude-latitude map of the radial component of the magnetic field in the HIRES region is presented in Figure ~\ref{model-path}. The HIRES computational domain extends about 88$^{\circ}$ in longitude, 48$^{\circ}$ in latitude, and up to 2.1 $\rm R_\odot$ from the Sun center. The models use variable grid spacing to achieve high spatial resolution in the lower corona (i.e., $0.002 R_{\sun}$) while covering a large coronal volume in and around the target region. Two paths, marked with blue curves, are selected according to the locations of the observed filaments. Path 1 represents the filament in the active region (F1), and the filament outside the active region (F2) is represented by Path 2. Next we modify the potential field to create cavities in the region above the selected paths, and then insert two thin flux bundles (representing the axial flux $\Phi_{\rm axi}$ of the flux rope) into the cavities. Circular loops are added around the flux bundle to represent the poloidal flux $F_{\rm pol}$ of the flux rope.  The resulting magnetic fields are not in force-free equilibrium. We then use the magneto-frictional relaxation to drive the field towards a force-free state \citep{ballegooijen2000, yang1986}. 

\begin{figure*}[!ht]
\centering\includegraphics[width=0.95\textwidth]{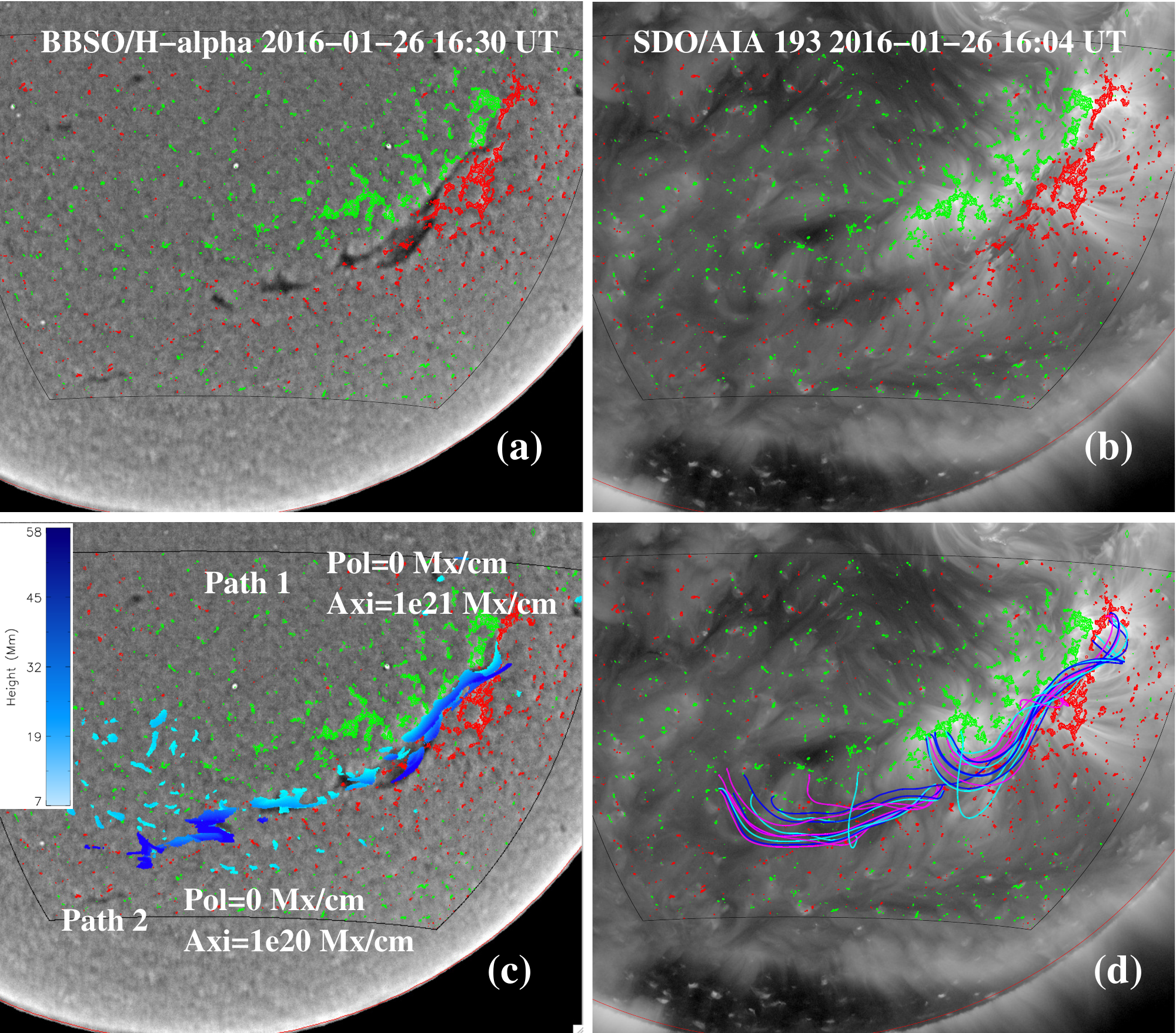}
\caption{Comparison between observations and the best-fit magnetic field model after a 30000-iteration relaxation. The left and right columns present BBSO/H${\alpha}$ and SDO/AIA 193\AA\ images. {The images in the bottom row are overlaid with field line dips (blue features in panel (c)) and selected field lines (colored lines in panel (d)) from the best-fit model. The blue color code in panel (c) indicates the heights of the position of dips in the field lines, i.e., sites where the field lines are locally horizontal and curved upward. The heights are measured in Mm from the solar surface}. Red and green contours represent positive and negative magnetic polarities {measured with} SDO/HMI.
\label{fig:model-obs}}
\end{figure*}

We construct a series of magnetic field models by varying the axial and poloidal fluxes of the inserted flux ropes. One of the best-fit models is presented in Figure ~\ref{fig:model-obs}. The dips of the models give the positions where the cool plasma reside. We have selected the best-fit model by matching the positions of the dips to the observed H${\alpha}$ filaments { (see Figs. \ref{fig:model-obs}(a) and (c)) and the EUV emission surrounding the filament channels as well (Figs. \ref{fig:model-obs}(b) and (d))}. The inserted flux bundles have the same poloidal flux of 0 Mx cm$^{-1}$, and the axial fluxes are 10$^{21}$ and 10$^{20}$ Mx for paths 1 and 2, respectively. The left and right columns present Big Bear Solar Observatory(BBSO)/H${\alpha}$ and SDO/AIA 193 {\AA} images. The images in the bottom row are overlaid with field line dips (Figs. \ref{fig:model-obs}(c)) and selected field lines (Figs. \ref{fig:model-obs}(d)) from the best-fit model. The dips of the model field lines match the observed H${\alpha}$ filaments very well. {We also see that the fields lines correspond very well to the dark EUV filament channel in Figure \ref{fig:model-obs}(d).} The filaments appear to be supported by two weakly twisted flux ropes or sheared arcades as represented by the model field lines.

\section{Characteristics of the prominence deduced from seismology}\label{sec:fieldgeometry}

In the H$\alpha$ analysis of the oscillation we have found a period decreasing from 69 $\pm$ 2 to 60 $\pm$ 2 minutes. In contrast, the analysis of the 171~\AA\ data shows a quite uniform periodicity of around one hour. However, our in depth study of the oscillation in S1 has revealed that the period after the initial stage is 54 minutes.  Thus, we find that that the periodicities in the system we analyze range from 54 $\pm$ 1 to 69 $\pm$ 2 minutes. According to the model of \citet{luna2012b,luna2012c,luna2016}, the period of the longitudinal fundamental mode is determined by the relation
\begin{equation}\label{eq:pendulum0}
\omega^2=\frac{4 \, \pi^2}{P^2} = \frac{g}{R} + \frac{4\, c_\mathrm{sc}^2}{l\, (L -l) \, \kappa}\, ,
\end{equation}
where $\omega$ is the angular frequency of the oscillation, $P$ its period, $g$ the solar gravity, $R$ the radius of curvature of the dipped field lines where the cool prominence resides, $L$ is the length of such field lines, $l$ is the length of the cool prominence thread, and $\kappa$ is the temperature contrast between the corona and the prominence (typically larger than 100). In Equation (\ref{eq:pendulum0}) there are two contributions to the oscillation associated with two restoring forces. The first term is associated with the gravity projected along the magnetic field lines. The second term is associated with the pressure gradients producing the slow mode oscillation. For typical prominence parameters the slow mode contribution is small \citep[see,][]{luna2012c,Zhang2013a}. Thus, for prominences the longitudinal oscillation period is determined by the curvature of the field lines as
\begin{equation}\label{eq:pendulum}
\omega^2=\frac{4 \, \pi^2}{P^2} \approx \frac{g}{R}\, ,
\end{equation}
indicating that the gravity projected along the field lines is the main restoring force. With this expression we can determine partially the geometry of the field lines, inferring the radii of curvature of the dips of the field lines $R$ from the period of the oscillation. Using the periods obtained in the time-distance analysis, $P= 54 - 69$ minutes and Equation (\ref{eq:pendulum}) we obtain a range of possible radii of curvature of $R= 73 \pm 3$ Mm and $R= 119 \pm 7$ Mm for the short and long periods respectively. {In Section \ref{sec:171-oscil-analysis} we speculated that the two phases of the temporal evolution of the oscillation are associated with change in the geometry of the field lines. In the first phase we have found a period of 69 minutes and in a second phase of 54 minutes. This could be related to a temporal evolution of the radius of curvature of the dips from 119 Mm to 73 Mm according to Equation (\ref{eq:pendulum}).}

In a prominence the magnetic structure must support the dense threads of plasma making up the prominence. The forces associated with the magnetic structure are the magnetic tension in the dipped part of the field lines and the magnetic pressure gradients. Both magnetic forces are almost in equilibrium with slight excess of the magnetic tension over magnetic pressure force that can provide support against gravity \citep{priest2014}. Thus, the magnetic tension in the dipped part of the tubes must be larger than the weight of the threads or equivalently
\begin{equation}\label{eq:force-balance}
\frac{B^2}{\mu \, R} -m\, n\, g \ge 0 \, ,
\end{equation}
where $B$ is the magnetic field strength at the bottom of the dip, $n$ is the {electron number} density in the thread, the mean particle mass $m = 1.27 \, m_p$ \citep{aschwanden2004}, and $m_p$ is the proton mass. Combining Equations (\ref{eq:pendulum}) and (\ref{eq:force-balance}) we obtain
\begin{equation}\label{eq:seismology-field}
B[\mathrm{G}] \ge 0.43 \left(\frac{n}{10^{11}\, \mathrm{cm}^{-3}}\right)^{1/2} \, P[\mathrm{min}] \,
\end{equation}
where $B$ is measured in Gauss, $n$ in $\mathrm{cm}^{-3}$ and $P$ in minutes. The electron density in this filament has not been measured, so we assume typical values in the range $10^{10} - 10^{11} \mathrm{cm}^{-3}$ \citep[see, e.g.,][]{labrosse2010}. We consider the range of typical values for electron density as the uncertainty of $n$. Considering the uncertainties of $n$ and the range of periods obtained in the analysis of the time-distance diagrams and their uncertainties, we obtain a minimum magnetic field strength range of $B\ge 7 - 30$G. The values obtained are consistent with typical directly measured field strengths \citep[see, e.g.,][]{mackay2010}.

\begin{figure}
\includegraphics[width=0.5\textwidth]{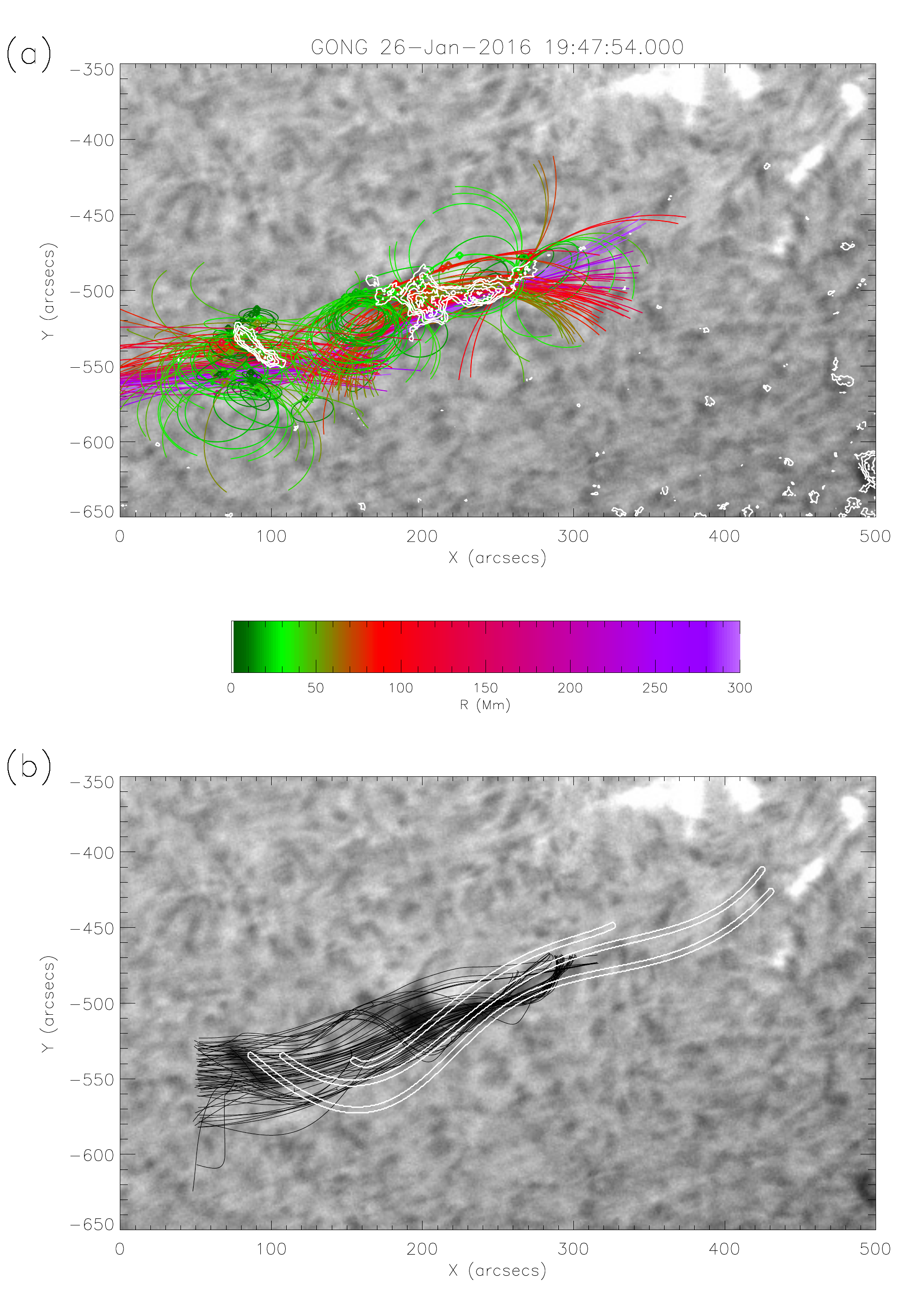}
\caption{(a) Plot of the region of interest in H$\alpha$ as in Figure \ref{fig:gong-data-map}. The colored lines are segments of the osculating circle tangent to the bottom positions of the dips. The segments plotted are a 200 arcsecs portion of the osculating circle diameter. Colors show the radius of curvature, $R$, of the osculating circle in Mm. 
We have over-plotted the H$\alpha$ intensity over the osculating circles in order to have a clear view of the position of the cool plasma (thin, white isocontour lines). (b) Plot of the dipped field lines (black curves) over the region of interest in H$\alpha$.  The slits S1--S3 are also over-plotted to compare the relative alignment of the magnetic field with the motion of the plasma.\label{fig:model17_curvature}}
\end{figure}

We compare the results of the large-amplitude seismology with the results of the reconstructed magnetic field model from Section \ref{sec:fieldgeometry}. We compute the positions of the bottoms of the dips in approximately 2000 field lines selected in a region close to F2. Once we have the positions of the dips we compute the radius of curvature of the field lines and the osculating circle \citep{struik1961} in these positions. The osculating circle is a circle with radius $R$ that fits the dipped part of the field lines. In Figure \ref{fig:model17_curvature}(a) a segment of the osculating circle is plotted for each dip. The segments shown in the figure correspond to 200 arcsecs centered at the bottom position of the dips. For dips with a small radius of curvature the plot shows the full circle whereas for large radii only a segment of the osculating circle is plotted. The colors represent the radius of curvature, $R$, of the osculating circles. The result obtained from seismology is that the radius of curvature is in the range between 73 and 119 Mm. This range of values corresponds to the red tones (see the color bar in the figure). {There is a} large bundle of osculating circles in red tones over the location of {the oscillating filament segment,} MF2. We find good agreement between the results of the geometry inferred from the seismology and the results of the field reconstruction. {There is also a bundle of osculating circles in red tones in SF2 indicating a similar geometry with the field of MF2.}

In general, the osculating circles are not vertical, forming an angle with respect to the vertical. Also in general the dipped part of the field lines is not a circular shape. In contrast, the geometry of the field is more complex, with curvatures depending on the position along the tube. Hence, depending on the amplitude of the displacement of the LALOs the relation between the period of the oscillation and the geometry can be more complex than the expression in Equation (\ref{eq:pendulum}). In addition, in expression (\ref{eq:pendulum}) it is assumed that the osculation circle is vertical, i.e., the dip is contained in a vertical plane. However, in realistic field-lines the inclination of the osculating circle produces a reduction of the projection of the gravity along the field lines or, equivalently, an increase of the effective radius of curvature. In a future theoretical study we will consider LALOs in non-uniform curvatures and inclined dips.

In Figure \ref{fig:model17_curvature}(b) {we have plotted as dark curves} a selection of field lines inferred from the field reconstruction. We have also plotted the artificial slits used to track the motion in 171~\AA\ data. We see that both the field lines and the artificial slits  are more or less aligned. There are some differences probably due to imperfect alignment of the slits with the threads and differences between the model and the actual magnetic field. Nevertheless, the alignment is quite good and consistent with the idea that the oscillation is aligned with the filament channel magnetic field.

\begin{figure}
\centering\includegraphics[width=0.5\textwidth]{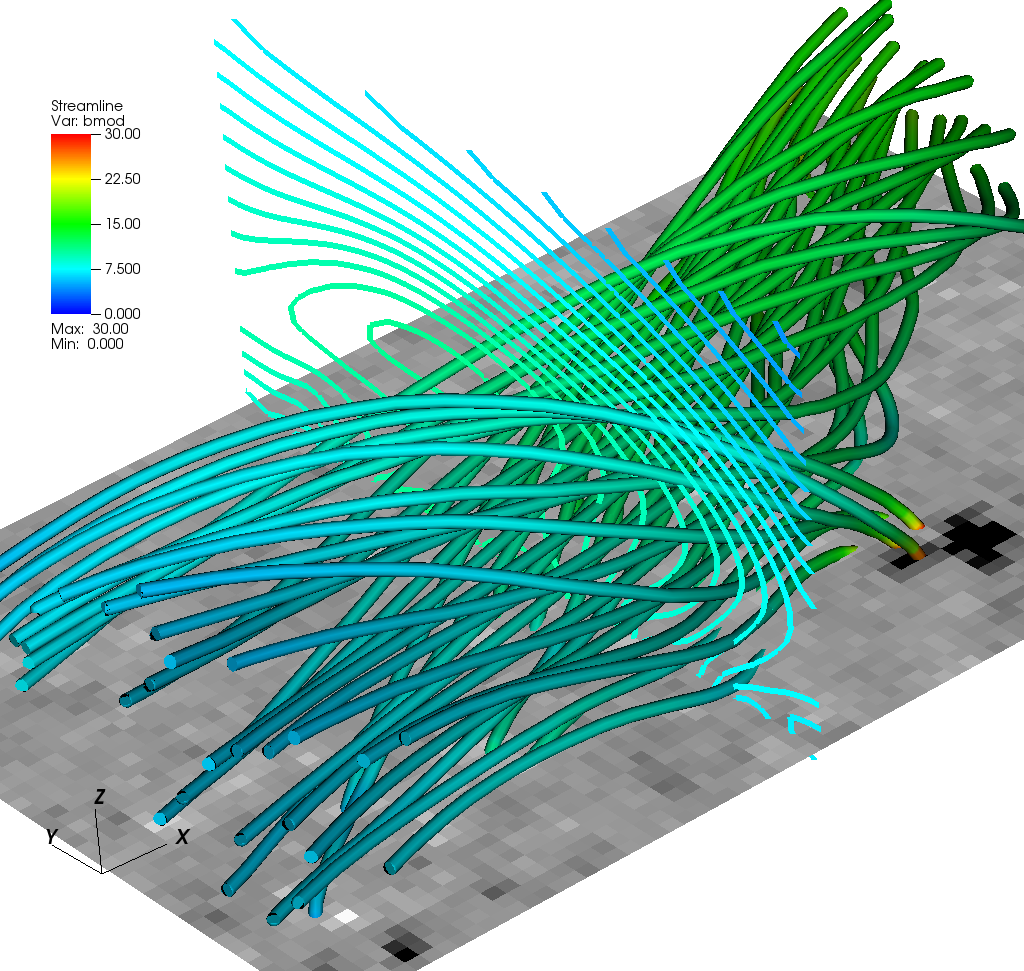}
\caption{Three dimensional view of the field lines used to compute the radius of curvature of the filament channel lines. The color of the lines indicates the magnetic field intensity. A vertical plane is also plotted in the north-south direction over the central position of F2. In this plane the contours of the field intensity are shown.
\label{fig:fieldstrenghth-3dview}}
\end{figure}

The reconstructed magnetic field model also provides the intensity of the magnetic field. In Figure \ref{fig:fieldstrenghth-3dview} we have plotted a 3D view of the field over the region of interest centered at F2. We also plot the magnetic field intensity in a vertical plane oriented in the north-south direction over the central position of F2. The color code indicates that in the region of the dipped parts of the field lines the magnetic field intensity is between 8 and 20 G. The values obtained are consistent with the lower limit obtained from seismology $\ge 7$ to 30 G. Note the range of values from seismology is very broad because of  large uncertainty in plasma density in Equation (\ref{eq:seismology-field}), and for some density values the lower limit could be larger than the values obtained from the field reconstruction. 

\section{Discussion and Conclusions}\label{sec:conclusions}

In this work we analyze an energetic large-amplitude longitudinal oscillation measured after the merging of two filaments on 2016 January 26. The initial configuration consists of two filaments, F1 and F2. F1 erupts and part of its plasma is merged with F2, triggering the oscillations in the central portion of F2. We have analyzed the oscillations with the time-distance technique using the GONG H$\alpha$ and SDO/AIA 171 \AA\ data. The time-distance analysis of the H$\alpha$ data reveals that the period of the oscillation changes with time from 69 $\pm$ 2 to 60 $\pm$ 2 minutes along the temporal sequence with a damping time of 85 $\pm$ 7 minutes. We have also constructed a time-distance diagram using the same slit but in the 171~\AA\ data. In 171 \AA\ it appears that the cool material consists of many threads visible as dark {absorbing} material surrounded by the {hotter} PCTR layer. These threads oscillate with apparently different phases and amplitudes. We have determined that the H$\alpha$ motion is the bulk or average motion of the fine-structure of the prominence. We have studied  the oscillations in the 171~\AA\ filter further by placing 3 curved slits defined manually to follow the flows. The time-distance diagrams reveal a relatively uniform  period between 60 $\pm$ 0.2 to 68 $\pm$ 0.4 minutes, depending on location. This suggests a quite uniform prominence structure. In one of the slits the velocity amplitude of the oscillations is 100 $\pm$ 7 $\kms$,  the largest ever reported. The detailed analysis of the oscillation in Slit 1 reveals that the oscillation shows two phases. In the first phase the period of the oscillation is around 69 minutes. In the second phase the period is reduced to 54 minutes, a value close to the value determined by \citep{zheng2017}. The measured damping in H$\alpha$ is almost double that obtained from 171~\AA\ data. The apparently stronger damping in H$\alpha$ might be due to a lower spatial resolution combined with the averaged motion of several threads moving out of phase.

Combining the results of the observations with the models of \citet{luna2012b} and \citet{luna2012c} we derive the characteristics of the field lines supporting the oscillating material, specifically the radius of curvature of the dips holding the prominence plasma (73 to 119 Mm),  and then, using common electron number density values for prominences, the magnetic field strength ($\ge 7$ to 30~G).

In order to make a comparison with these values, we constructed the magnetic field background using the flux-rope insertion method. The resulting magnetic field is a flux rope with a small poloidal field and a strong axial field, indicating that prominence field is strongly sheared. The characteristics of the flux rope derived by the MHD reconstruction are consistent with the seismology results: the radii of curvature of the model dips are in the range of 70 Mm to 130 Mm and the magnetic field strength is 8 to 20 G. This is the first time these two approaches have been compared. {In the future it would also be interesting to compare results like ours to still other sources of information concerning the magnetic field including constraints derived from spectro-polarimetric or radio observations.}

This work shows that prominence seismology is a powerful tool to determine the filament structure. In future works  models for large-amplitude oscillations will be improved by doing three-dimensional numerical experiments. In addition, it is necessary to understand the influence on the oscillation of non-uniform curvature and inclined dips. More studies of the triggering processes leading to LALOs are also important, both to better understand the oscillations and to shed light on processes associated with the triggers, including reconnection and related solar activity.

\acknowledgements

Wavelet software was provided by C. Torrence and G. Compo, and is available at URL: \emph{http://paos.colorado.edu/research/wavelets/}. SDO is a mission of NASA\rq{}s Living With a Star Program. AIA and HMI data are courtesy of the NASA/\textit{SDO} science teams. The Global Oscillation Network Group (GONG) Program is managed by the NSO and operated by AURA, Inc. under a cooperative agreement with the NSF. The data are acquired by instruments operated by the Big Bear Solar Observatory, High Altitude Observatory, Learmonth Solar Observatory, Udaipur Solar Observatory, Instituto de Astrof\'isica de Canarias, and Cerro Tololo Interamerican Observatory. The operation of Big Bear Solar Observatory is supported by NJIT, US NSF AGS-1250818, and NASA NNX13AG14G grants.

This work was initiated during the first International Space Science Institute (ISSI) team meeting in Bern managed by Nicolas Labrosse on ``Solving the Prominence Paradox.'' M. Luna also acknowledges ISSI support to Team 314, ``Large-Amplitude Oscillations in Solar Prominences'' led by this author. M. Luna acknowledges the support by the Spanish Ministry of Economy and Competitiveness through project AYA2014-55078-P. Y.~Su is supported by the One Hundred Talent Program of CAS and NSFC 11473071, as well as the Youth Fund of JiangSu No. BK20141043. R. Chandra wants to thank the Observatoire de Paris for financial support during his one month visit. T.~Kucera was supported by the NASA Heliophysics Guest Investigator Program.


\begin{thebibliography}

\bibitem[{{Anzer} \& {Heinzel}(2005)}]{anzer2005}
{Anzer}, U., \& {Heinzel}, P. 2005, \apj, 622, 714

\bibitem[{{Arregui} {et~al.}(2012){Arregui}, {Oliver}, \&
  {Ballester}}]{arregui2012}
{Arregui}, I., {Oliver}, R., \& {Ballester}, J.~L. 2012, Living Reviews in
  Solar Physics, 9, 2

\bibitem[{Aschwanden(2004)}]{aschwanden2004}
Aschwanden, M.~J. 2004, Physics of the Solar Corona. An Introduction

\bibitem[{{Aulanier} {et~al.}(1998){Aulanier}, {Demoulin}, {van
  Driel-Gesztelyi}, {Mein}, \& {Deforest}}]{aulanier1998}
{Aulanier}, G., {Demoulin}, P., {van Driel-Gesztelyi}, L., {Mein}, P., \&
  {Deforest}, C. 1998, \aap, 335, 309

\bibitem[{{Aulanier} {et~al.}(2006){Aulanier}, {DeVore}, \&
  {Antiochos}}]{aulanier2006}
{Aulanier}, G., {DeVore}, C.~R., \& {Antiochos}, S.~K. 2006, \apj, 646, 1349

\bibitem[{Ballester {et~al.}(2016)Ballester, Carbonell, Soler, \&
  Terradas}]{ballester2016}
Ballester, J.~L., Carbonell, M., Soler, R., \& Terradas, J. 2016, Astronomy and
  Astrophysics

\bibitem[{{Berger} {et~al.}(2008){Berger}, {Shine}, {Slater}, {Tarbell},
  {Title}, {Okamoto}, {Ichimoto}, {Katsukawa}, {Suematsu}, {Tsuneta}, {Lites},
  \& {Shimizu}}]{berger2008}
{Berger}, T.~E., {Shine}, R.~A., {Slater}, G.~L., {et~al.} 2008, \apjl, 676,
  L89

\bibitem[{Berger {et~al.}(2010)Berger, Slater, Hurlburt, Shine, Tarbell, Title,
  Lites, Okamoto, Ichimoto, Katsukawa, Magara, Suematsu, \&
  Shimizu}]{berger2010}
Berger, T.~E., Slater, G., Hurlburt, N., {et~al.} 2010, The Astrophysical
  Journal, 716, 1288

\bibitem[{Bi {et~al.}(2014)Bi, Jiang, Yang, Hong, Li, Yang, \& Yang}]{bi2014}
Bi, Y., Jiang, Y., Yang, J., {et~al.} 2014, The Astrophysical Journal, 790, 100

\bibitem[{Bobra {et~al.}(2008)Bobra, van Ballegooijen, \& DeLuca}]{bobra2008}
Bobra, M.~G., van Ballegooijen, A.~A., \& DeLuca, E.~E. 2008, The Astrophysical
  Journal, 672, 1209

\bibitem[{Carbonell \& Ballester(1991)}]{carbonell1991}
Carbonell, M., \& Ballester, J.~L. 1991, Astronomy and Astrophysics, 249, 295

\bibitem[{{Chandra} {et~al.}(2011){Chandra}, {Schmieder}, {Mandrini},
  {D{\'e}moulin}, {Pariat}, {T{\"o}r{\"o}k}, \& {Uddin}}]{Chandra11}
{Chandra}, R., {Schmieder}, B., {Mandrini}, C.~H., {et~al.} 2011, \solphys,
  269, 83

\bibitem[{Chen {et~al.}(2008)Chen, Innes, \& Solanki}]{chen2008}
Chen, P.~F., Innes, D.~E., \& Solanki, S.~K. 2008, Astronomy and Astrophysics,
  484, 487

\bibitem[{Dud{\'\i}k {et~al.}(2012)Dud{\'\i}k, Aulanier, Schmieder, Zapi{\'o}r,
  \& Heinzel}]{dudik2012}
Dud{\'\i}k, J., Aulanier, G., Schmieder, B., Zapi{\'o}r, M., \& Heinzel, P.
  2012, The Astrophysical Journal, 761, 9

\bibitem[{{E.~Tandberg-Hanssen}(1995)}]{tandberg1995}
{E.~Tandberg-Hanssen}, ed. 1995, Astrophysics and Space Science Library, Vol.
  199, {The nature of solar prominences}

\bibitem[{{Engvold}(2015)}]{engvold15}
{Engvold}, O. 2015, in Astrophysics and Space Science Library, Vol. 415, Solar
  Prominences, ed. J.-C. {Vial} \& O.~{Engvold}, 31

\bibitem[{Eto {et~al.}(2002)Eto, Isobe, Narukage, Asai, Morimoto, Thompson,
  Yashiro, Wang, Kitai, Kurokawa, \& Shibata}]{eto2002}
Eto, S., Isobe, H., Narukage, N., {et~al.} 2002, Publications of the
  Astronomical Society of Japan, 54, 481

\bibitem[{{Filippov}(2015)}]{Filippov16}
{Filippov}, B. 2015, \mnras, 453, 1550

\bibitem[{Gilbert {et~al.}(2008)Gilbert, Daou, Young, Tripathi, \&
  Alexander}]{gilbert2008}
Gilbert, H.~R., Daou, A.~G., Young, D., Tripathi, D., \& Alexander, D. 2008,
  \apj, 685, 629

\bibitem[{Gunar \& Mackay(2016)}]{gunar2016}
Gunar, S., \& Mackay, D.~H. 2016, Astronomy and Astrophysics, 592, A60

\bibitem[{Gunar {et~al.}(2013)Gunar, Mackay, Anzer, \& Heinzel}]{gunar2013}
Gunar, S., Mackay, D.~H., Anzer, U., \& Heinzel, P. 2013, Astronomy and
  Astrophysics, 551, A3

\bibitem[{Gunar {et~al.}(2014)Gunar, Schwartz, Dud{\'\i}k, Schmieder, Heinzel,
  \& Jur{\v c}{\'a}k}]{gunar2014}
Gunar, S., Schwartz, P., Dud{\'\i}k, J., {et~al.} 2014, Astronomy and
  Astrophysics, 567, A123

\bibitem[{{Heinzel} {et~al.}(2008){Heinzel}, {Schmieder}, {F{\'a}rn{\'{\i}}k},
  {Schwartz}, {Labrosse}, {Kotr{\v c}}, {Anzer}, {Molodij}, {Berlicki},
  {DeLuca}, {Golub}, {Watanabe}, \& {Berger}}]{Heinzel2008a}
{Heinzel}, P., {Schmieder}, B., {F{\'a}rn{\'{\i}}k}, F., {et~al.} 2008, \apj,
  686, 1383

\bibitem[{Hershaw {et~al.}(2011)Hershaw, Foullon, Nakariakov, \&
  Verwichte}]{hershaw2011}
Hershaw, J., Foullon, C., Nakariakov, V.~M., \& Verwichte, E. 2011, Astronomy
  and Astrophysics, 531, A53

\bibitem[{{Hyder}(1966)}]{hyder1966}
{Hyder}, C.~L. 1966, \zap, 63, 78

\bibitem[{Isobe \& Tripathi(2006)}]{isobe2006}
Isobe, H., \& Tripathi, D. 2006, Astronomy and Astrophysics, 449, L17

\bibitem[{Isobe {et~al.}(2007)Isobe, Tripathi, Asai, \& Jain}]{isobe2007}
Isobe, H., Tripathi, D., Asai, A., \& Jain, R. 2007, Solar Physics, 246, 89

\bibitem[{{Jibben} {et~al.}(2016){Jibben}, {Reeves}, \& {Su}}]{jibben2016}
{Jibben}, P., {Reeves}, K., \& {Su}, Y. 2016, Frontiers in Astronomy and Space
  Sciences, 3, 10

\bibitem[{{Jing} {et~al.}(2006){Jing}, {Lee}, {Spirock}, \& {Wang}}]{jing2006}
{Jing}, J., {Lee}, J., {Spirock}, T.~J., \& {Wang}, H. 2006, \solphys, 236, 97

\bibitem[{{Jing} {et~al.}(2003){Jing}, {Lee}, {Spirock}, {Xu}, {Wang}, \&
  {Choe}}]{jing2003}
{Jing}, J., {Lee}, J., {Spirock}, T.~J., {et~al.} 2003, \apjl, 584, L103

\bibitem[{{Labrosse} {et~al.}(2010){Labrosse}, {Heinzel}, {Vial}, {Kucera},
  {Parenti}, {Gun{\'a}r}, {Schmieder}, \& {Kilper}}]{labrosse2010}
{Labrosse}, N., {Heinzel}, P., {Vial}, J., {et~al.} 2010, \ssr, 151, 243

\bibitem[{{Lemen} {et~al.}(2012){Lemen}, {Title}, {Akin}, {Boerner}, {Chou},
  {Drake}, {Duncan}, {Edwards}, {Friedlaender}, {Heyman}, {Hurlburt}, {Katz},
  {Kushner}, {Levay}, {Lindgren}, {Mathur}, {McFeaters}, {Mitchell}, {Rehse},
  {Schrijver}, {Springer}, {Stern}, {Tarbell}, {Wuelser}, {Wolfson}, {Yanari},
  {Bookbinder}, {Cheimets}, {Caldwell}, {Deluca}, {Gates}, {Golub}, {Park},
  {Podgorski}, {Bush}, {Scherrer}, {Gummin}, {Smith}, {Auker}, {Jerram},
  {Pool}, {Soufli}, {Windt}, {Beardsley}, {Clapp}, {Lang}, \&
  {Waltham}}]{Lemen2012a}
{Lemen}, J.~R., {Title}, A.~M., {Akin}, D.~J., {et~al.} 2012, \solphys, 275, 17

\bibitem[{{Leroy} {et~al.}(1983){Leroy}, {Bommier}, \&
  {Sahal-Brechot}}]{Leroy1983a}
{Leroy}, J.~L., {Bommier}, V., \& {Sahal-Brechot}, S. 1983, \solphys, 83, 135

\bibitem[{{Leroy} {et~al.}(1984){Leroy}, {Bommier}, \&
  {Sahal-Brechot}}]{Leroy1984a}
---. 1984, \aap, 131, 33

\bibitem[{{Li} \& {Zhang}(2012)}]{Li2012}
{Li}, T., \& {Zhang}, J. 2012, \apjl, 760, L10

\bibitem[{Liu {et~al.}(2013)Liu, Liu, Xu, Liu, Kliem, \& Wang}]{liu2013}
Liu, R., Liu, C., Xu, Y., {et~al.} 2013, The Astrophysical Journal, 773, 166

\bibitem[{Lomb(1976)}]{lomb1976}
Lomb, N.~R. 1976, Astrophysics and Space Science, 39, 447

\bibitem[{{Luna} {et~al.}(2012){Luna}, {D{\'{\i}}az}, \& {Karpen}}]{luna2012c}
{Luna}, M., {D{\'{\i}}az}, A.~J., \& {Karpen}, J. 2012, \apj, 757, 98

\bibitem[{Luna {et~al.}(2016{\natexlab{a}})Luna, D{\'\i}az, Oliver, Terradas,
  \& Karpen}]{luna2016}
Luna, M., D{\'\i}az, A.~J., Oliver, R., Terradas, J., \& Karpen, J.
  2016{\natexlab{a}}, Astronomy and Astrophysics, 593, A64

\bibitem[{{Luna} \& {Karpen}(2012)}]{luna2012b}
{Luna}, M., \& {Karpen}, J. 2012, \apjl, 750, L1

\bibitem[{Luna {et~al.}(2014)Luna, Knizhnik, Muglach, Karpen, Gilbert, Kucera,
  \& Uritsky}]{luna2014}
Luna, M., Knizhnik, K., Muglach, K., {et~al.} 2014, The Astrophysical Journal,
  785, 79

\bibitem[{Luna {et~al.}(2016{\natexlab{b}})Luna, Terradas, Khomenko, Collados,
  \& Vicente}]{luna2016a}
Luna, M., Terradas, J., Khomenko, E., Collados, M., \& Vicente, A.~d.
  2016{\natexlab{b}}, The Astrophysical Journal, 817, 157

\bibitem[{{Mackay} {et~al.}(2010){Mackay}, {Karpen}, {Ballester}, {Schmieder},
  \& {Aulanier}}]{mackay2010}
{Mackay}, D.~H., {Karpen}, J.~T., {Ballester}, J.~L., {Schmieder}, B., \&
  {Aulanier}, G. 2010, \ssr, 151, 333

\bibitem[{{Martin}(1998)}]{martin98}
{Martin}, S.~F. 1998, \solphys, 182, 107

\bibitem[{Moreton \& Ramsey(1960)}]{moreton1960}
Moreton, G.~E., \& Ramsey, H.~E. 1960, Publications of the Astronomical Society
  of the Pacific, 72, 357

\bibitem[{{Morgan} \& {Druckm{\"u}ller}(2014)}]{Morgan14}
{Morgan}, H., \& {Druckm{\"u}ller}, M. 2014, \solphys, 289, 2945

\bibitem[{Okamoto {et~al.}(2004)Okamoto, Nakai, Keiyama, Narukage, Ueno, Kitai,
  Kurokawa, \& Shibata}]{okamoto2004}
Okamoto, T.~J., Nakai, H., Keiyama, A., {et~al.} 2004, \apj, 608, 1124

\bibitem[{Okamoto {et~al.}(2007)Okamoto, Tsuneta, Berger, Ichimoto, Katsukawa,
  Lites, Nagata, Shibata, Shimizu, Shine, Suematsu, Tarbell, \&
  Title}]{okamoto2007}
Okamoto, T.~J., Tsuneta, S., Berger, T.~E., {et~al.} 2007, Science, 318, 1577

\bibitem[{{Oliver} \& {Ballester}(2002)}]{oliver2002}
{Oliver}, R., \& {Ballester}, J.~L. 2002, \solphys, 206, 45

\bibitem[{{Parenti} {et~al.}(2012){Parenti}, {Schmieder}, {Heinzel}, \&
  {Golub}}]{Parenti2012}
{Parenti}, S., {Schmieder}, B., {Heinzel}, P., \& {Golub}, L. 2012, \apj, 754,
  66

\bibitem[{{Pesnell} {et~al.}(2012){Pesnell}, {Thompson}, \&
  {Chamberlin}}]{Pesnell12}
{Pesnell}, W.~D., {Thompson}, B.~J., \& {Chamberlin}, P.~C. 2012, \solphys,
  275, 3

\bibitem[{Pouget {et~al.}(2006)Pouget, Bocchialini, \& Solomon}]{pouget2006}
Pouget, G., Bocchialini, K., \& Solomon, J. 2006, SOHO-17. 10 Years of SOHO and
  Beyond, 617, 141

\bibitem[{{Priest}(2014)}]{priest2014}
{Priest}, E. 2014, {Magnetohydrodynamics of the Sun}

\bibitem[{{Ramsey} \& {Smith}(1966)}]{ramsey1966}
{Ramsey}, H.~E., \& {Smith}, S.~F. 1966, \aj, 71, 197

\bibitem[{Ruderman \& Luna(2016)}]{ruderman2016}
Ruderman, M.~S., \& Luna, M. 2016, Astronomy and Astrophysics, 591, A131

\bibitem[{{Scherrer} {et~al.}(2012){Scherrer}, {Schou}, {Bush}, {Kosovichev},
  {Bogart}, {Hoeksema}, {Liu}, {Duvall}, {Zhao}, {Title}, {Schrijver},
  {Tarbell}, \& {Tomczyk}}]{Scherrer12}
{Scherrer}, P.~H., {Schou}, J., {Bush}, R.~I., {et~al.} 2012, \solphys, 275,
  207

\bibitem[{Schmieder {et~al.}(2013)Schmieder, Kucera, Knizhnik, Luna,
  Lopez-Ariste, \& Toot}]{schmieder2013}
Schmieder, B., Kucera, T.~A., Knizhnik, K., {et~al.} 2013, The Astrophysical
  Journal, 777, 108

\bibitem[{{Schmieder} {et~al.}(2004){Schmieder}, {Lin}, {Heinzel}, \&
  {Schwartz}}]{schmieder2004a}
{Schmieder}, B., {Lin}, Y., {Heinzel}, P., \& {Schwartz}, P. 2004, \solphys,
  221, 297

\bibitem[{Schmieder {et~al.}(2014)Schmieder, Roudier, Mein, Mein, Malherbe, \&
  Chandra}]{schmieder2014}
Schmieder, B., Roudier, T., Mein, N., {et~al.} 2014, Astronomy and
  Astrophysics, 564, 104

\bibitem[{{Struik}(1961)}]{struik1961}
{Struik}, D.~J. 1961, {Lectures on Classical Differential Geometry, by Dirk J.
  Struik, New York, USA: Dover Publications Inc., 1961}

\bibitem[{Su {et~al.}(2011)Su, Surges, van Ballegooijen, Deluca, \&
  Golub}]{su2011}
Su, Y., Surges, V., van Ballegooijen, A., Deluca, E., \& Golub, L. 2011, The
  Astrophysical Journal, 734, 53

\bibitem[{Su \& van Ballegooijen(2012)}]{su2012b}
Su, Y., \& van Ballegooijen, A. 2012, The Astrophysical Journal, 757, 168

\bibitem[{{Su} {et~al.}(2015){Su}, {van Ballegooijen}, {McCauley}, {Ji},
  {Reeves}, \& {DeLuca}}]{su2015}
{Su}, Y., {van Ballegooijen}, A., {McCauley}, P., {et~al.} 2015, \apj, 807, 144

\bibitem[{Su {et~al.}(2009)Su, van Ballegooijen, Schmieder, Berlicki, Guo,
  Golub, \& Huang}]{su2009}
Su, Y., van Ballegooijen, A., Schmieder, B., {et~al.} 2009, The Astrophysical
  Journal, 704, 341

\bibitem[{{T{\"o}r{\"o}k} {et~al.}(2011){T{\"o}r{\"o}k}, {Chandra}, {Pariat},
  {D{\'e}moulin}, {Schmieder}, {Aulanier}, {Linton}, \& {Mandrini}}]{Torok11}
{T{\"o}r{\"o}k}, T., {Chandra}, R., {Pariat}, E., {et~al.} 2011, \apj, 728, 65

\bibitem[{Torrence \& Compo(1998)}]{torrence1998}
Torrence, C., \& Compo, G.~P. 1998, Bulletin of the American Meteorological
  Society, 79, 61

\bibitem[{{Tripathi} {et~al.}(2009){Tripathi}, {Isobe}, \&
  {Jain}}]{tripathi2009}
{Tripathi}, D., {Isobe}, H., \& {Jain}, R. 2009, \ssr, 149, 283

\bibitem[{{van Ballegooijen}(2004)}]{vanballegooijen2004}
{van Ballegooijen}, A.~A. 2004, \apj, 612, 519

\bibitem[{van Ballegooijen {et~al.}(2000)van Ballegooijen, Priest, \&
  Mackay}]{ballegooijen2000}
van Ballegooijen, A.~A., Priest, E.~R., \& Mackay, D.~H. 2000, The
  Astrophysical Journal, 539, 983

\bibitem[{{Vr{\v s}nak} {et~al.}(2007){Vr{\v s}nak}, {Veronig}, {Thalmann}, \&
  {{\v Z}ic}}]{vrsnak2007}
{Vr{\v s}nak}, B., {Veronig}, A.~M., {Thalmann}, J.~K., \& {{\v Z}ic}, T. 2007,
  \aap, 471, 295

\bibitem[{Yang {et~al.}(1986)Yang, Sturrock, \& Antiochos}]{yang1986}
Yang, W.~H., Sturrock, P.~A., \& Antiochos, S.~K. 1986, The Astrophysical
  Journal, 309, 383

\bibitem[{{Zhang} {et~al.}(2012){Zhang}, {Chen}, {Xia}, \&
  {Keppens}}]{zhang2012}
{Zhang}, Q.~M., {Chen}, P.~F., {Xia}, C., \& {Keppens}, R. 2012, \aap, 542, A52

\bibitem[{{Zhang} {et~al.}(2013){Zhang}, {Chen}, {Xia}, {Keppens}, \&
  {Ji}}]{Zhang2013a}
{Zhang}, Q.~M., {Chen}, P.~F., {Xia}, C., {Keppens}, R., \& {Ji}, H.~S. 2013,
  \aap, 554, A124

\bibitem[{Zhang {et~al.}(2017)Zhang, Li, Zheng, Su, \& Ji}]{zhang2017}
Zhang, Q.~M., Li, T., Zheng, R.~S., Su, Y.~N., \& Ji, H.~S. 2017, The
  Astrophysical Journal, 842, 27

\bibitem[{Zheng {et~al.}(2017)Zheng, Zhang, Chen, Wang, Du, Li, \&
  Yang}]{zheng2017}
Zheng, R., Zhang, Q., Chen, Y., {et~al.} 2017, The Astrophysical Journal, 836,
  160

\bibitem[{{Zirker} {et~al.}(1998){Zirker}, {Engvold}, \& {Martin}}]{zirker1998}
{Zirker}, J.~B., {Engvold}, O., \& {Martin}, S.~F. 1998, \nat, 396, 440

\bibitem[{{Zirker} {et~al.}(1994){Zirker}, {Engvold}, \& {Yi}}]{zirker1994}
{Zirker}, J.~B., {Engvold}, O., \& {Yi}, Z. 1994, \solphys, 150, 81

\end{thebibliography}

\end{document}